\documentclass[12pt,a4paper]{article} 
\pdfoutput=1
\usepackage[whole]{bxcjkjatype} 
\usepackage{amsmath}
\usepackage{amssymb}
\usepackage{amsthm}

\usepackage{setspace}
\usepackage{empheq}
\usepackage{bm}
\usepackage{color}
\usepackage{cite} 
\usepackage{braket}
\usepackage{float}

\theoremstyle{definition}
\newtheorem{dfn}{Definition}

\usepackage{hyperref}
\usepackage{graphicx}

\catcode `@=11 \@addtoreset{equation}{section} \catcode `@=12

\def\Tr{{\rm Tr}}

\begin{document}
\setlength{\oddsidemargin}{0cm}

\begin{titlepage}

    \begin{center}
        {\LARGE	
        Bootstrapping Shape Invariance: \\
Numerical Bootstrap as a Detector of Solvable Systems
		}
    \end{center}
    \vspace{1.2cm}
    \baselineskip 18pt 
    \renewcommand{\thefootnote}{\fnsymbol{footnote}}

    \begin{center}

		Yu {\sc Aikawa}$^{a}$\footnote{%
		E-mail address: aikawa.yu.17(at)shizuoka.ac.jp} and
        Takeshi {\sc Morita}$^{a, b}$\footnote{%
            E-mail address: morita.takeshi(at)shizuoka.ac.jp
        }
        
        \renewcommand{\thefootnote}{\arabic{footnote}}
        \setcounter{footnote}{0}
        
        \vspace{0.4cm}
        
        {\it
		a. Graduate School of Science and Technology, Shizuoka University\\
		836 Ohya, Suruga-ku, Shizuoka 422-8529, Japan
            \vspace{0.2cm}
            \\
			b. Department of Physics,
            Shizuoka University \\
            836 Ohya, Suruga-ku, Shizuoka 422-8529, Japan 
        }

    \end{center}

    
    \vspace{1.5cm}
    
    \begin{abstract}
		Determining the solvability of a given quantum mechanical system is generally challenging. We discuss that the numerical bootstrap method can help us to solve this question in one-dimensional quantum mechanics. We show that the bootstrap method can derive exact energy eigenvalues in systems with shape invariance, which is a sufficient condition for solvability and which many solvable systems satisfy. The information of the annihilation operators is also obtained naturally, and thus the bootstrap method tells us why the system is solvable. We numerically demonstrate this explicitly for shape invariant potentials: harmonic oscillators, Morse potentials, Rosen-Morse potentials and hyperbolic Scarf potentials.  Therefore, the numerical bootstrap method can determine the solvability of a given unknown system if it satisfies shape invariance.
	\end{abstract}

    
\end{titlepage}

\tableofcontents

\section{Introduction}
\label{sec-Intro}

Numerical analysis is indispensable for the development of modern physics. However, it is difficult to obtain exact results even for solvable systems using ordinary numerical methods such as the Monte Carlo method. Also, determining by ordinary numerical analysis whether a given system is solvable or not is hard.
Since solvability is crucial information for a system, it is quite significant if the solvability of the system could be determined by numerical analysis. 

On the other hand, the numerical bootstrap method, which has been used in conformal field theories in recent years, is known to be able to obtain {\em exact} bounds (e.g., exact upper and lower bounds) on the allowed values of the physical quantities \cite{Poland:2018epd}. Here, the exact bound means that, the physical quantities cannot take the values outside the bound absolutely. The regions not excluded by the bounds are called ``allowed regions", and the quantities can take values only in these regions. This is one of the features of the numerical bootstrap method that distinguishes it from existing numerical analyses. 

The bootstrap method has also been applied to quantum mechanics by Han et al \cite{Han:2020bkb}. Using this method, one can obtain isolated allowed regions for observables such as the energy and the expectation value $\langle x^n \rangle$ corresponding to the energy eigenstates in one-dimensional quantum mechanical systems. The accuracy of this method is controlled by the size of the bootstrap matrix, which we will review in Sec.~\ref{sec-review-bootstrap}. For a sufficiently large matrix size, the size of the isolated allowed region can be very small, and we obtain the value of the observable with high precision. 
Thus, the bootstrap method is a powerful tool for evaluating observables in quantum mechanics. (Not only that, the bootstrap method is related to an extension of the uncertainty relation \cite{Morita:2022zuy}, and this is conceptually interesting because it implies a direct relationship between the uncertainty relation and the energy eigenvalues. See footnote \ref{ftnt-XP} for more details.)

The numerical bootstrap method in quantum mechanics has been applied to various models \cite{Berenstein:2021dyf, Bhattacharya:2021btd, Aikawa:2021eai, Berenstein:2021loy, Tchoumakov:2021mnh, Aikawa:2021qbl, Du:2021hfw, Nakayama:2022ahr, Li:2022prn, Khan:2022uyz, Berenstein:2022ygg, Blacker:2022szo, Berenstein:2022unr, Berenstein:2023ppj, Sword:2024gvv, Li:2024rod, Lawrence:2024mnj, Berenstein:2025itw}. Among them, interesting results have been obtained in harmonic oscillators \cite{Aikawa:2021qbl} and P\"{o}schl-Teller potentials \cite{Sword:2024gvv}, both of which are known to be solvable. (In this paper, we call a system ``solvable", if the exact energy eigenvalues of the system can be derived. We call a system solvable even if not all energy eigenvalues are obtained.) In these solvable models, even at a small matrix size, isolated allowed regions are given by single points, and the energies at these points precisely match the known energy eigenvalues. 
We call such points ``allowed points".
Thus, the numerical bootstrap method can reproduce exact solutions, and this is a significant difference from ordinary numerical methods. These results suggest that the bootstrap method is useful as a tool for numerically finding solvable systems. In other words, if the bootstrap method leads to exact energy eigenvalues through the allowed points, it means that the system is solvable.

It should be emphasized that the allowed points obtained by the numerical bootstrap method are not always strictly points due to the limitation of the numerical analysis.
The allowed points are slightly smeared due to numerical errors\footnote{The borders of the allowed regions (the exact upper and lower bounds) mentioned above may also be smeared due to numerical error. However the numerical error is typically much smaller than the error bar of other numerical method, and, for this reason, the bounds are regarded as ``exact", and they are often used to test the validity of the results of other numerical analyses.  }. Let us call such numerically obtained approximated allowed points ``smeared allowed points" to distinguish them from the allowed points, which are strictly points.
However, for a fixed size of the bootstrap matrix, the sizes of the isolated allowed regions in non-solvable models are finite, and we can easily distinguish them from the smeared allowed points in solvable systems. 
The purpose of this paper is not to obtain the strictly exact solutions, but to establish the numerical method that can examine whether the system is solvable. (We also introduce a bootstrap analysis using the characteristic polynomial, which provides us the strictly exact solutions (allowed points) analytically. However, this method does not always work.)

However, it is not clear whether the bootstrap method is really useful for finding solvable systems. One possibility is that the bootstrap method can only find solvability for very limited models such as harmonic oscillators and P\"{o}schl-Teller potentials, but cannot find it for other solvable systems. To understand this problem, we focus on shape invariance \cite{Gendenshtein:1983skv} and study its relationship with the bootstrap method.

Shape invariance is one of the sufficient conditions for solvability. In fact, many solvable systems such as harmonic oscillators, P\"{o}schl-Teller potentials, Coulomb potentials, Morse potentials and so on satisfy shape invariance. (See Refs.~\cite{Cooper:1994eh, Sasaki:2014tka, Nasuda:2024cod} for reviews of these topics.) In this paper, we analytically show that the bootstrap method can derive exact solutions, if the systems satisfy shape invariance. Thus, exact solutions for a large number of solvable systems can be obtained by the bootstrap method. In addition, the information of the annihilation operators is also obtained naturally in the bootstrap method. Therefore, the bootstrap method tells us why the system is solvable.

As examples, we analyze Morse potentials \cite{PhysRev.34.57}, Rosen-Morse potentials \cite{PhysRev.42.210} and hyperbolic Scarf potentials \cite{JWDabrowska_1988, Levai:1989eaa, Alvarez-Castillo:2006uzq}, which satisfy shape invariance and are solvable. We show that the numerical bootstrap method yields exact solutions which reproduces the known analytic solutions. \\

In general, it is a non-trivial problem to determine whether a given system satisfies shape invariance or not, and it is also generally difficult to determine whether the system is solvable or not. 
We will discuss that the numerical bootstrap method can determine the solvability of a system at least if it is shape invariant.
(We also show one class of systems, in which the system does not satisfy shape invariance but the bootstrap method can derive exact solutions in Appendix \ref{app-DT}.)
However, there may be situations where the bootstrap method cannot provide exact solutions even if the system is solvable, and obtaining exact solutions in the bootstrap method is a sufficient condition for solvability.\\

The organization of this paper is as follows.
In Sec.~\ref{sec-solvable}, we review the numerical bootstrap method and show that the bootstrap method can derive exact solutions for some solvable models.
Then, we propose that the bootstrap method can be used to determine whether a given system is solvable in Sec.~\ref{sec-proposal}.
To strengthen our proposal, we show that the bootstrap method can derive the exact solutions for shape invariant systems in Sec.~\ref{sec-SI-bootstrap}. (The review of shape invariance is given in Sec.~\ref{sec-SI}.)
In Sec.~\ref{sec-examples-bootstrap}, we analyze harmonic oscillators and Morse potentials in more detail by using the bootstrap method, and see how the bootstrap method works in shape invariant systems explicitly. 
We summarize our results and discuss future prospects in Sec.~\ref{sec-discussion}.
In Appendix \ref{app-bootstrap-details}, we provide several useful equations for our bootstrap analysis.
In Appendix \ref{app-numerical}, some details of the numerical computations including a sample code are given.
In Appendix \ref{app-Morse}, we show the details of the analysis of the Morse potentials.
In Appendix \ref{app-SI-bootstrap}, Rosen-Morse potentials and hyperbolic Scarf potentials are investigated by the bootstrap method as examples of shape invariant systems.
In Appendix \ref{app-DT}, we show an example of a system that does not satisfy shape invariance but the bootstrap method can derive exact solutions.\\

We will use the units $\hbar=1$ in this paper.

\section{Solvable systems in bootstrap method}
\label{sec-solvable}

In this section, we provide an introduction to the numerical bootstrap method \cite{Han:2020bkb}. In particular, we demonstrate that the bootstrap method can derive exact solutions in solvable systems. We also discuss several features of solvable systems in the bootstrap method. For instance, the annihilation operators are naturally derived in the bootstrap method in the case of harmonic oscillators.

\subsection{Review of bootstrap method}
\label{sec-review-bootstrap}

We briefly review the bootstrap method in an one-dimensional quantum mechanical system with a Hamiltonian $H=H(x,p)$ \cite{Han:2020bkb}.
The idea of the bootstrap method is deriving the spectrum of the system from the positivities of some selected observables.
Suppose that we take $K$ well-defined operators $\{ O_n \}$, ($n=1,\cdots,K$). For example $O_1=x$, $O_2=p$ and so on.
Then we define the following operator from them,
\begin{align}
	\tilde{O}=  \sum_{n=1}^{K} c_{n} O_n,
	\label{ops-sample}
\end{align}
where $\{ c_{n} \}$ are some constants.
Since $ \langle \alpha | O^\dagger O  | \alpha  \rangle \ge 0$ is satisfied for any well-defined state $|\alpha \rangle $ and any well-defined operator $O$ in the system, 
\begin{align}
	\langle \alpha | \tilde{O}^\dagger \tilde{O} | \alpha  \rangle \ge 0
	\label{positive-cond}
\end{align}
is satisfied too for arbitrary constants $\{ c_{n} \}$.
Hence, the following $K\times K$ matrix ${\mathcal M}$ has to be positive-semidefinite \cite{Han:2020bkb}\footnote{${\mathcal M} \succeq 0$ denotes that the matrix ${\mathcal M}$ is ``positive-semidefinite", i.e. all the eigenvalues of ${\mathcal M}$ are non-negative.},
\begin{align}
	{\mathcal M}:=
	\begin{pmatrix}
		\left\langle O_1^\dagger O_1 \right\rangle & \left\langle O_1^\dagger O_2 \right\rangle     & \cdots & \left\langle O_1^\dagger O_K \right\rangle   \\
		\left\langle O_2^\dagger O_1 \right\rangle &  \left\langle O_2^\dagger O_2 \right\rangle &  \cdots & \left\langle O_2^\dagger O_K \right\rangle   \\
			\vdots & \vdots   & \ddots & \vdots  \\
			\left\langle O_K^\dagger O_1 \right\rangle
		& \left\langle O_K^\dagger O_2 \right\rangle
		& \cdots & \left\langle O_K^\dagger O_K \right\rangle
	\end{pmatrix}
	\succeq 0.
	\label{bootstrap}
\end{align}
Here we have omitted $\alpha$ in $|\alpha \rangle $.
This strongly constrains possible expectation values of the operators.
We call ${\mathcal M}$ as a bootstrap matrix.
Note that, as $K$ increases, the constraint would become stronger.

From now, we focus on an energy eigenstate with an energy eigenvalue $E$,
and we take $| \alpha \rangle = | E \rangle  $.
Then, the energy eigenstate has to satisfy the following two additional constraints,
\begin{align}
	&\langle E| \left[ H, O \right] | E  \rangle =0 ,
	\label{HO=0} \\
	&\langle E| HO | E \rangle =E \langle E| O | E  \rangle=\langle E| OH | E \rangle ,
	\label{HO=EO}
\end{align}
for any well-defined operator $O$\footnote{
In one-dimensional quantum mechanics, if $x$ is the operator on a half-line \cite{Berenstein:2022ygg} or on an interval \cite{Sword:2024gvv}, the conditions \eqref{HO=0} and \eqref{HO=EO} have to be modified due to anomalies \cite{Esteve:2002nt}.}.
We survey the allowed values of $E$ that are consistent with these constraints and the condition ${\mathcal M} \succeq 0$.
If the constraints are sufficiently strong, the ranges of $E$ that satisfy the constraints are highly restricted and may become point-like corresponding to the energy eigenvalues. (These ranges are called ``allowed regions".)
In this way, we may obtain the energy eigenvalues by the bootstrap method.

\subsection{Examples: Harmonic oscillator vs. Anharmonic oscillator}
\label{sec-HO-AHO}
We illustrate the bootstrap method using two examples: the harmonic oscillator and the anharmonic oscillator,
\begin{align}
	H=\frac{1}{2}p^2+\frac{1}{2}x^2
	\label{H-HO} ,\\
	H=\frac{1}{2}p^2+\frac{1}{2}x^2+\frac{1}{4}x^4.
	\label{H-AHO}
	\end{align}
In particular, since we are interested in the solvability of the bootstrap method, we pay attention to how the results differ between the solvable (the harmonic oscillator) and non-solvable (the anharmonic oscillator) cases.

To use the bootstrap method, we first need to choose a set of operators $\{O_n\}$ in Eq.~\eqref{ops-sample} and construct the bootstrap matrix \eqref{bootstrap}. In fact, the results depend on the choice of operators. The details have been studied in Ref.~\cite{Aikawa:2021qbl}, and a part of our discussion in this section is based on that work.

As a natural set of operators, we take $\{x^m p^n\}$, since the Hamiltonian \eqref{H-HO} and \eqref{H-AHO} are described by these operators. Then, we define the operator,
\begin{align}
	\tilde{O}_{xp}:=  \sum_{m=0}^{K_x} \sum_{n=0}^{K_p} c_{mn} x^m p^n.
	\label{operators-XP}
\end{align}
where $ c_{mn} $ are constants correspond to $c_n$ in Eq.~\eqref{ops-sample}, and the integers $K_x$ and $K_p$ determine the size of the bootstrap matrix as $K= (K_x+1)(K_p+1) $. By using this $\tilde{O}_{xp}$, we construct the bootstrap matrix \eqref{bootstrap}\footnote{\label{ftnt-XP}
	We can show that the condition 
\begin{align*}
		\begin{pmatrix}
			1 & \left\langle x \right\rangle  & \left\langle p \right\rangle  \\
			\left\langle x	\right\rangle & \left\langle x^2	\right\rangle & \left\langle xp \right\rangle \\
			\left\langle p	\right\rangle  &  \left\langle px \right\rangle & \left\langle p^2 \right\rangle   
		\end{pmatrix}
		\succeq 0,
	\end{align*}
is equivalent to the uncertainty relation $ \Braket{\Delta x^2} \Braket{\Delta p^2}   \ge \hbar^2 /4 $ \cite{Morita:2022zuy, Simon:1997es, Curtright:2001jn}.
	Thus the condition ${\mathcal M}_{xp}\succeq 0 $, which involves higher moment operators $\{x^m p^n\}$, can be interpreted as an extension of the uncertainty relation.
	},
\begin{align}
	{\mathcal M}_{xp}:=
	\begin{pmatrix}
		1 & \left\langle x \right\rangle  & \left\langle p \right\rangle   & \cdots  \\
		\left\langle x	\right\rangle & \left\langle x^2	\right\rangle & \left\langle xp \right\rangle &  \cdots  \\
		\left\langle p	\right\rangle  &  \left\langle px \right\rangle & \left\langle p^2 \right\rangle  & \cdots   \\
		\vdots & \vdots  & \vdots  & \ddots \\
	\end{pmatrix}.
	\label{bootstrap-XP}
\end{align}
Here, we have assumed that the energy eigenstate is normalized as $\langle E | E \rangle =1$.
Each element of this bootstrap matrix takes the form $\Bra{E} p^k x^l p^m \ket{E}$. (Here we have explicitly written $\ket{E}$ to emphasize that the state is an energy eigenstate, but we usually omit it in this paper.) 
Now we impose the conditions \eqref{HO=0} and \eqref{HO=EO}.
We substitute $O=x^a p^b$ into these conditions where $a$ and $b$ are some non-negative integers, and, after some computations, they are reduced to the following three equations \cite{Han:2020bkb, Aikawa:2021qbl},
\begin{align}
	&	n(n-1)(n-2) \langle x^{n-3} \rangle -8n \langle x^{n-1} V(x)\rangle +8 n E \langle x^{n-1}  \rangle -4 \langle x^{n} V'(x)\rangle=0, \nonumber \\
	&	\langle x^{m} p \rangle  = \frac{im}{2}  \langle x^{m-1} \rangle, \nonumber \\
&	\langle x^{m} p^{n+2}\rangle  = m(m-1)\langle x^{m-2} p^{n}\rangle +  2 i m  \langle x^{m-1} p^{n+1}\rangle +   2E \langle x^{m
	} p^{n} \rangle -2 \langle x^{m} V(x) p^{n} \rangle.
	\label{eq-xp-recurrence}
\end{align}
Here $m$ and $n$ are non-negative integers and we have set $H=p^2/2+V(x)$ assuming that $V(x)$ is a polynomial potential like Eqs.~\eqref{H-HO} and \eqref{H-AHO}.
The first equation relates $\langle x^n \rangle$ with different power $n$, and the second and third equations relate $\langle x^m p^n \rangle$ with lower $n$.
Using the commutation relation $[x,p]=i$ and solving these equations recurrently (usually by computer), the matrix element can be expressed in the form,
\begin{align} 
	\Bra{E}p^k x^l p^m \Ket{E} = \sum_{0 \le n \le N_{\rm max}} C^{klm}_{n}(E) \Bra{E} x^n \Ket{E}.
	\label{eq-pxp-x}  
\end{align}
Here $C^{klm}_{N}(E)$ is a polynomial of $E$ and $N_{\rm max}$ is an integer, and they are determined once the potential $V(x)$ is given.

In the harmonic oscillator \eqref{H-HO}, $N_{\rm max}=0$ and all the matrix elements are expressed by $E$.
In the case of the anharmonic oscillator \eqref{H-AHO}, $N_{\rm max}=2$ and the elements are expressed by $E$ and $\Braket{x}$ and $\Braket{x^2}$\footnote{$\Braket{x}=0$ if we impose the parity. However, the bootstrap method correctly derives energy eigenvalues without using such additional conditions from the beginning \cite{Morita:2022zuy}. Indeed, the linear programming analysis for $\Braket{x}$ provides correct energy eigenvalues as demonstrated in Sec.\ref{sec-SDP}. 
}. Then, the bootstrap matrix \eqref{bootstrap-XP} becomes \cite{Aikawa:2021qbl}
\begin{align}
	{\mathcal M}_{xp}^{(HO)}:=
	\begin{pmatrix}
		1 & 0  & 0   & \cdots  \\
		0 &  E & \frac{i}{2} &  \cdots  \\
		0  &  -\frac{i}{2} & E  & \cdots   \\
		\vdots & \vdots  & \vdots  & \ddots \\
	\end{pmatrix}, \quad
	{\mathcal M}_{xp}^{(AHO)}:=
	\begin{pmatrix}
		1 & \left\langle x	\right\rangle  & 0   & \cdots  \\
		\left\langle x	\right\rangle  &  \left\langle x^2	\right\rangle & \frac{i}{2} &  \cdots  \\
		0  &  -\frac{i}{2} & \frac{1}{3}\left( 4E- \left\langle x^2	\right\rangle \right)  & \cdots   \\
		\vdots & \vdots  & \vdots  & \ddots \\
	\end{pmatrix}.
	\label{bootstrap-XP2}
\end{align}
(Some useful equations to obtain these expressions for the bootstrap matrices are summarized in Appendix \ref{app-bootstrap-details}.)
We survey the allowed regions in which these bootstrap matrices become positive-semidefinite. For this purpose, linear programming is typically employed. Evaluating the characteristic polynomial of the bootstrap matrix is also useful. We review these two methods in the next subsections and apply them to obtain the energy eigenvalues of the harmonic and the anharmonic oscillators.

\subsubsection{Linear Programming}
\label{sec-SDP}

As can be seen from Eq.~\eqref{eq-pxp-x}, each component of the bootstrap matrix is a polynomial in $E$ but linear in $\Braket{x^n}$ ($n=1,\cdots,N_{\rm max}$). 
In such a case, numerical linear programming is applicable.
We select one of an operator $\Braket{x^m} \in  \{ \Braket{x^n} \}_{n=1,\cdots,N_{\rm max}}$, and, by fixing the value of $E$, we can efficiently determine the upper and lower bounds of  $ \Braket{x^m}$ that satisfy ${\mathcal M} \succeq 0$ using numerical linear programming. The results obtained by numerical linear programming are not strictly exact, but they are highly accurate. If ${\mathcal M} $ cannot be positive-semidefinite for any value of $\Braket{x^m}$, the value of $E$ is excluded from the allowed region. 
By repeating this procedure, we obtain the allowed region for ($E$, $\Braket{x^m}$) that satisfies ${\mathcal M} \succeq 0$. If the allowed region shrinks to a point-like region as the size of the bootstrap matrix increases, we obtain the energy and the expectation value $\Braket{x^m}$ corresponding to the energy eigenstate, where the width of the allowed region can be regarded as an error bar.
If we are interested in other observable $ \Braket{x^k} \in  \{ \Braket{x^n} \}_{n=1,\cdots,N_{\rm max}}$ ($k \neq m$), we can repeat the same procedure with respect to $ \Braket{x^k}$. Note that the allowed region for $E$ obtained from $\Braket{x^k}$ should agree with that obtained from $\Braket{x^m}$ in numerical linear programming, except for numerical error.

\begin{figure}[t]
	\centering
	\includegraphics[scale=0.7]{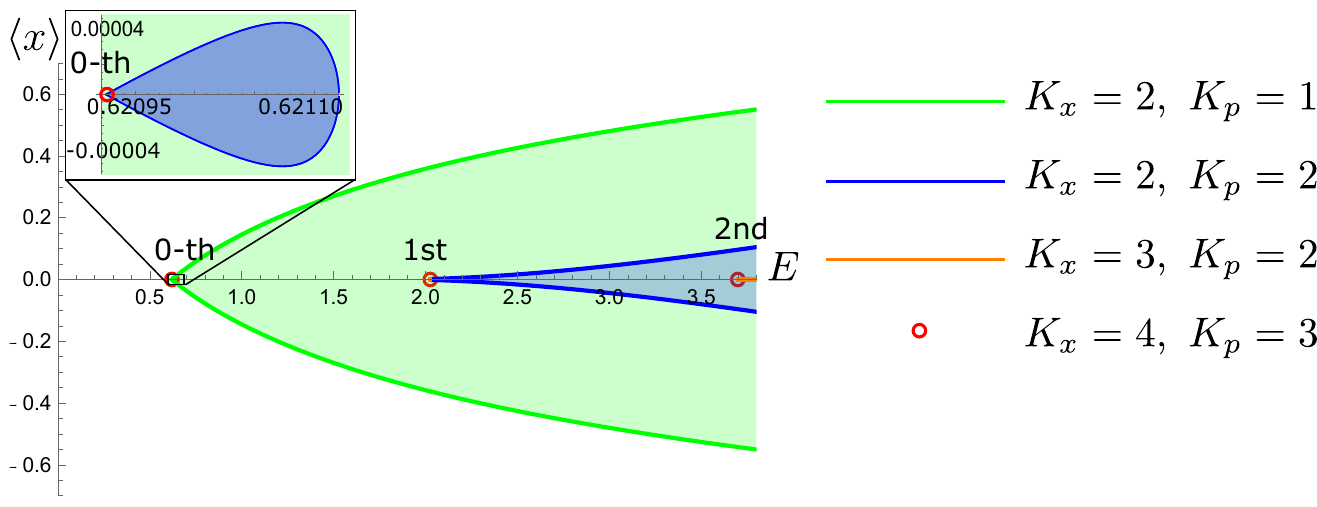}
	\caption{Numerical bootstrap result for the anharmonic oscillator \eqref{H-AHO}.
	We used the bootstrap matrix ${\mathcal M}_{xp}^{(AHO)}$ \eqref{bootstrap-XP2} and $K_x$ and $K_p$ defined in Eq.~\eqref{operators-XP} determine the size of the bootstrap matrix.
	We use the numerical linear programming to find the upper and lower bounds of $\Braket{x}$ at each fixed value of $E$.
	The colored regions show the allowed regions in the $(E,\Braket{x})$ plane that satisfy the condition ${\mathcal M}_{xp}^{(AHO)} \succeq 0$. The results for $(K_x,K_p)=(4,3)$ are almost points, and they are highlighted with red circles. 
	The regions for	$(K_x,K_p)=(3,2)$ and $(K_x,K_p)=(4,3)$ are difficult to see in this figure, and see Table \ref{Table-AHO-XP} for the allowed values of $E$. 
	As $K_x$ and $K_p$ increase, the allowed regions shrink, and converge to the results at $(K_x,K_p)=(4,3)$. Here, $\Braket{x}$ converges to 0, as expected from the parity.}
	\label{fig-x4-x}
\end{figure}

\begin{figure}[t]
	\centering
	\includegraphics[scale=0.7]{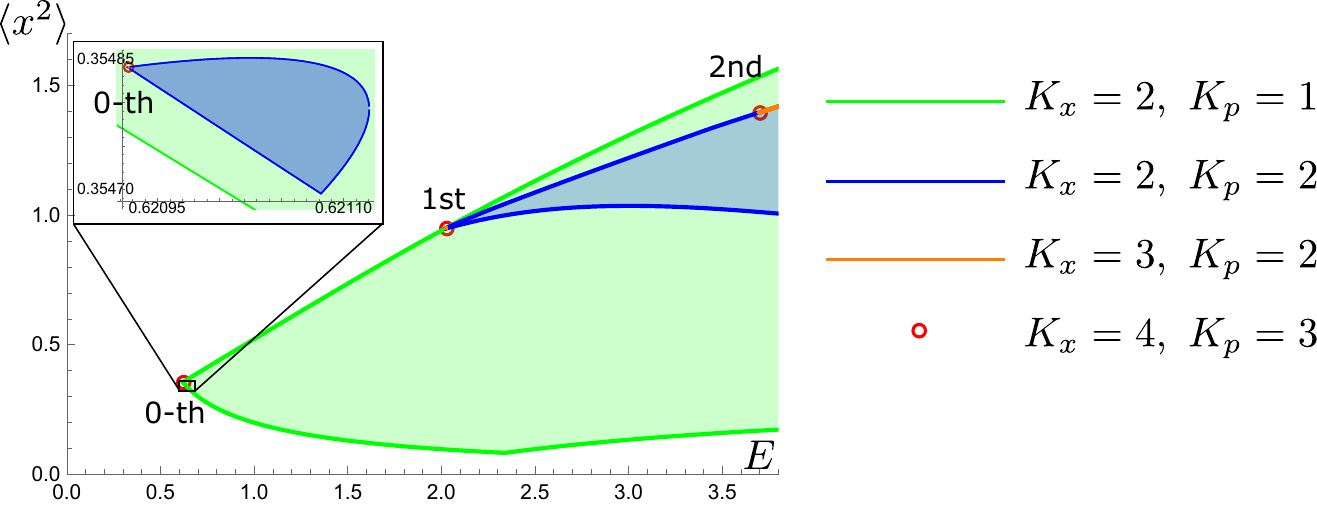}
	\caption{	
	Numerical bootstrap result for the anharmonic oscillator \eqref{H-AHO}.
	We use the same analysis as the one in Fig.~\ref{fig-x4-x}, but we solve the linear programming with respect to $\Braket{x^2}$ instead of $\Braket{x}$.
	The obtained allowed regions for $E$ is consistent with Fig.~\ref{fig-x4-x}, and they are summarized in Table \ref{Table-AHO-XP}. Again, the regions for $(K_x,K_p)=(3,2)$ and $(K_x,K_p)=(4,3)$ are difficult to see in this figure, and see Table \ref{Table-AHO-XP} for the details.} 
	\label{fig-x4}
\end{figure}

Let us apply this method to the anharmonic oscillator. 
The bootstrap matrix \eqref{bootstrap-XP2} is linear in $\Braket{x}$ and $\Braket{x^2}$, and we perform a numerical linear programming for these two variables at each fixed value of $E$.
We use the Mathematica package ``SemidefiniteOptimization" and take ``Mosek" for the ``Method" option, when we solve the linear programming in this paper (Our Mathematica is version 14.2 and a sample code is given in Appendix \ref{app-sample-code}).
The results for $\Braket{x}$ and $\Braket{x^2}$ are shown in Fig.~\ref{fig-x4-x} and \ref{fig-x4}, respectively.
In these two figures, we see that the allowed regions of $E$ are isolated when the size of the bootstrap matrix is large. As the size increases, the allowed region shrinks to a point-like region.
The allowed values of $E$ obtained from $\Braket{x}$ and $\Braket{x^2}$ are almost the same, and they are summarized in Table~\ref{Table-AHO-XP}.
For example, at $(K_x,K_p)=(3,2)$, the ground state energy is restricted to $0.62092702 \le E  \le 0.62092706$. Thus, $E$ is obtained with the accuracy of $10^{-8}$. In this way, the energy eigenvalues can be obtained with high precision. \\

In the case of the harmonic oscillator, since the components of the bootstrap matrix \eqref{bootstrap-XP2} are polynomials of $E$, and we cannot use linear programming. Instead of it, we investigate the characteristic polynomial of the bootstrap matrix \eqref{bootstrap-XP2} as we explain in the next subsection.

\begin{table}[t]
	\centering
	\begin{tabular}{|c||c|c|c|}
		\hline
		& $n=0$   & $n=1$  & $n=2$   \\ 
		\hline
		\hline
		$(K_x,K_p)=(1,1)$  &      \multicolumn{3}{c |}{$0.43 \le E$}    \\
		\hline 
		$(K_x,K_p)=(1,2)$  &      \multicolumn{3}{c |}{$0.49 \le E$}    \\
\hline 		$(K_x,K_p)=(2,1)$   &      \multicolumn{3}{c |}{$0.62 \le E$}    \\
\hline 
		$(K_x,K_p)=(2,2)$   &   $0.620926 \le E  \le 0.621121$ &    \multicolumn{2}{c |}{$2.026\le E$}    \\
\hline 
$(K_x,K_p)=(3,2)$  &   $0.62092702 \le E  \le 0.62092706
$ &  $2.02597 \le E \le 2.02601 $ & $3.698\le E$    \\
\hline 
	$(K_x,K_p)=(4,3)$ & $E= 0.6209270298...$ & $E=2.0259661641...$ & $E$=3.6984503...   \\
		\hline 
	\end{tabular}
	\caption{
		Energy spectrum of the anharmonic oscillator \eqref{H-AHO} via the bootstrap method. 
		This data corresponds to the allowed regions shown in Fig.~\ref{fig-x4}, and the results in Fig.~\ref{fig-x4-x} are consistent with this table.
		 The allowed values of $E$ converge as the matrix size increases, and are almost point-like for $(K_x,K_p)=(4,3)$. However, these results cannot be considered exact solutions because they are not single points but very small isolated allowed regions.
		}
	\label{Table-AHO-XP}
\end{table}

\subsubsection{Derivation of the exact allowed region via characteristic polynomial}
\label{sec-SDP-char}

The allowed region satisfying ${\mathcal M} \succeq 0$ can also be derived by investigating the characteristic polynomial of the bootstrap matrix ${\mathcal M}$. At the boundary of the allowed region, at least one of the eigenvalues of ${\mathcal M}$ becomes zero. 
To find the boundary, we consider the characteristic polynomial of ${\mathcal M}$,
\begin{align} 
  \det({\mathcal M} - \lambda I) = 0,
  \label{eq-char-poly}
\end{align}
and define the coefficient of the lowest power of $\lambda$ as $f_{\mathcal M}(E,\{\braket{x^n}\}_{n=1,\cdots,N_{\rm max}})$.
Since $f_{\mathcal M}$ is the product of all the eigenvalues of ${\mathcal M}$ except the eigenvalues that are identically zero, it becomes zero at the boundary of the allowed region.
Thus, $E$ and $\{\braket{x^n}\}_{n=1,\cdots,N_{\rm max}}$ satisfying
\begin{align} 
  f_{\mathcal M}(E,\{\braket{x^n}\}_{n=1,\cdots,N_{\rm max}})=0
  \label{eq-bootstrap-char}
 \end{align}
are the candidates for the boundary of the allowed region.
Once we find such $E$ and $\{\braket{x^n}\}_{n=1,\cdots,N_{\rm max}}$, we can check whether they are indeed on the boundary by examining the eigenvalues of ${\mathcal M}$ in the vicinity of these points.

Let us apply this method to the harmonic oscillator \eqref{H-HO}.
As an example, we construct the bootstrap matrix from the operators $\{O_m\}:=\{1,x,p,xp,x^2\}$,
\begin{align}
	{\mathcal M}=
	\begin{pmatrix}
		1 & \left\langle x \right\rangle  & \left\langle p \right\rangle   &  \left\langle xp \right\rangle &  \left\langle x^2 \right\rangle  \\
		\left\langle x	\right\rangle & \left\langle x^2	\right\rangle & \left\langle xp \right\rangle &  \left\langle x^2p \right\rangle &  \left\langle x^3 \right\rangle   \\
		\left\langle p	\right\rangle  &  \left\langle px \right\rangle & \left\langle p^2 \right\rangle  &  \left\langle pxp \right\rangle &  \left\langle px^2 \right\rangle    \\
		\left\langle px	\right\rangle  &  \left\langle px^2 \right\rangle & \left\langle p x p \right\rangle  &  \left\langle px^2p \right\rangle &  \left\langle px^3 \right\rangle    \\
		\left\langle x^2	\right\rangle & \left\langle x^3	\right\rangle & \left\langle x^2p \right\rangle &  \left\langle x^3p \right\rangle &  \left\langle x^4 \right\rangle   \\
	\end{pmatrix}=
	\begin{pmatrix}
		1 & 0  & 0   &  \frac{i}{2} & E  \\
		0 & E &  \frac{i}{2} & 0 &  0   \\
		0  & - \frac{i}{2} & E  & 0 & 0    \\
		-  \frac{i}{2}  &  0 & 0  &  \frac{5}{8} + \frac{1}{2}E^2  &   -\frac{3i}{2}E    \\
		E & 0 & 0 &   \frac{3i}{2}E &   \frac{3}{8} + \frac{3}{2}E^2   \\
	\end{pmatrix}
	,
	\label{bootstrap-HO}
\end{align}
where we have used Eqs.~\eqref{eq-xp-ordering} and \eqref{eq-recurrence-HO}.
By computing the characteristic polynomial \eqref{eq-char-poly} of this matrix, we find that the lowest power of $\lambda$ is 0 and obtain
\begin{align} 
	  f_{\mathcal M}(E)=
\det {\mathcal M}= \frac{1}{4} \left(E-\frac{1}{2}\right)^2\left(E+\frac{1}{2}\right)^2 \left(E-\frac{3}{2}\right)\left(E+\frac{3}{2}\right).
\label{eq-bootstrap-char-HO}
 \end{align}
Thus, $f_{\mathcal M}$ becomes zero at $E=\pm 1/2$, $\pm 3/2$, and these are the candidates for the boundaries of the allowed regions. By evaluating the eigenvalues of ${\mathcal M}$ around these points, we find that the allowed regions are given by $E=1/2$ or $E\ge 3/2$.
Importantly, we obtain a single point allowed region at $E=1/2$ which precisely corresponds to the ground state energy of the harmonic oscillator.

We can repeat this procedure for larger bootstrap matrices. The computation becomes more complicated, but we can use the function \texttt{Solve} in Mathematica, which can solve the algebraic equation $f_{\mathcal M}=0$ exactly.
The results are summarized in Table \ref{Table-HO-XP}.
The bootstrap method precisely reproduces the energy eigenvalues of the harmonic oscillator $E=n+1/2$, ($n=0,1,2,\cdots$).
Thus, solving the condition \eqref{eq-bootstrap-char} is a powerful method to obtain the exact energy eigenvalues. 
(We will explain why the bootstrap method can reproduce the exact results in Sec.~\ref{sec-SI-bootstrap-energy}.)\\

For a general polynomial potential, $f_{\mathcal M}$ is a multi-variable function of $E$ and $\{\Braket{x^m} \}_{n=1,\cdots,N_{\rm max}}$, and $f_{\mathcal M}=0$ defines multi-dimensional surfaces.
In solvable systems, these surfaces may become point sets as in the case of the harmonic oscillator, and the values of the observables at these points may coincide with the exact solutions $\{E,\Braket{x},\Braket{x^2},\cdots\}=\{E,\Braket{x}_n,\Braket{x^2}_n,\cdots\}$ ($n=0,1,2,\cdots$) at the $n$-th energy eigenstate (where $\Braket{x^m}_n $ is the expectation value of $x^m$ for the $n$-th energy eigenstate).
We call such points as ``allowed points", and finding them is crucial to obtain the exact result.

To derive allowed points in the multi-variable case, the condition $f_{\mathcal M}=0$ may be insufficient.
At allowed points, the following conditions should also hold,
\begin{align} 
 0= \frac{\partial f_{\mathcal M}}{ \partial E} = \frac{\partial f_{\mathcal M}}{ \partial \braket{x^1 }} = \cdots = \frac{\partial f_{\mathcal M}}{ \partial \braket{x^{N_{\rm max}} }}  .
  \label{eq-bootstrap-char-deriv}
 \end{align}
(Indeed, in the harmonic oscillator case, $\frac{\partial f_{\mathcal M}}{ \partial E}=0$ is satisfied at the point $E=1/2$ from Eq.~\eqref{eq-bootstrap-char-HO}.) 
The number of these conditions is equal to the number of the variables  $E$ and $\{\Braket{x^m} \}_{n=1,\cdots,N_{\rm max}}$, and these equations are sufficient to determine the values of these variables.
Again, the solutions of these equations are just candidates for the allowed points or the boundaries of the allowed regions, and we have to check it by evaluating the eigenvalues of ${\mathcal M}$ around these points or solving numerical linear programming discussed in the previous subsection.
(We will apply this method to the Morse potential and the Rosen-Morse potential and obtain the exact solutions. See, for example, Table \ref{Table-Morse-XP}.)

Therefore, investigating the characteristic polynomial of the bootstrap matrix is a powerful method for solvable systems. However, when the size of the bootstrap matrix is large or when there are many independent variables $\{\Braket{x^m} \}_{n=1,\cdots,N_{\rm max}}$, it may take a long time to solve Eqs.~\eqref{eq-bootstrap-char} and \eqref{eq-bootstrap-char-deriv} by Mathematica (or it may stop and fail to return any solution). Thus, if the exact solutions are not required, the linear programming method described in Sec.~\ref{sec-SDP} is more convenient\footnote{If we solve (a part of) Eqs.~\eqref{eq-bootstrap-char} and \eqref{eq-bootstrap-char-deriv} in unsolvable systems, we may precisely obtain the upper and lower bounds of the isolated allowed regions as solutions of these algebraic equations. But such solutions may not be required, since the bounds move as the size of the bootstrap matrix changes and the importance of these solutions is unclear.}.

\begin{table}[t]
	\centering
	\begin{tabular}{|c||c|c|c|c|c|}
		\hline
		& $n= 0 $  & $n=1  $ & $n= 2$  & $n= 3$ & $n= 4$ \\ 
		\hline
		\hline
		 $(K_x,K_p)=(1,1)$ &   \multicolumn{5}{c |}{$1/2 \le E $}\\
		\hline 
		$(K_x,K_p)=(2,1)$ &  1/2 & \multicolumn{4}{c |}{$3/2 \le E $}\\
		\hline 
		$(K_x,K_p)=(3,1)$ &  1/2 &  3/2 & \multicolumn{3}{c |}{$5/2 \le E $}\\
		\hline 
		$(K_x,K_p)=(2,2)$ &  1/2 &  3/2 & \multicolumn{3}{c |}{$5/2 \le E $}\\
		\hline 
		$(K_x,K_p)=(3,2)$ &  1/2 &  3/2 & 5/2 & \multicolumn{2}{c |}{$7/2 \le E $}\\
		\hline 
		$(K_x,K_p)=(3,3)$ &  1/2 &  3/2 & 5/2 & 7/2  & $9/2 \le E $\\
		\hline 
		exact & 1/2 & 3/2 & 5/2 &  7/2 &  9/2   \\
		\hline 
	\end{tabular}
	\caption{Energy spectrum of the harmonic oscillator \eqref{H-HO} for the first five eigenstates ($n=0, \cdots , 4$) obtained by solving the equation \eqref{eq-bootstrap-char}. The bootstrap matrix constructed from the operators \eqref{operators-XP} is used, and the exact results are correctly derived. 
	}
	\label{Table-HO-XP}
\end{table}

\subsubsection{Solvable vs. Non-solvable systems}
\label{sec-solvable-nonsolvable}

We have seen that, in both the harmonic oscillator and anharmonic oscillator cases, better results are obtained by increasing the size $K_x$ and $K_p$, indicating that the size is a parameter to improve the accuracy of the bootstrap method. However, the response to increasing the size is quite different between the two cases. In the case of the harmonic oscillator (Table \ref{Table-HO-XP}), the ground state is first obtained exactly, and as the size is increased, the exact solutions are obtained from the lower energy level. Thus, the number of the exact solutions obtained increases as the size increases.

In the case of the anharmonic oscillator (Table \ref{Table-AHO-XP}), on the other hand, the number of the isolated regions corresponding to the energy eigenstates increases with the size. Each region also shrinks as the size increases. Thus, with increasing the size, the number of the energy eigenvalues obtained increases and the accuracy for each eigenstate is also improved.

Therefore, the results obtained by the bootstrap method are qualitatively completely different between the harmonic and anharmonic oscillators. This suggests that the numerical bootstrap method may be useful as a tool for a detection of solvable systems.

\subsubsection{Caution: Operator dependence}
\label{sec-HO-caution}

In the case of the harmonic oscillator, the exact energy eigenvalues can be obtained by the numerical bootstrap method. However, the bootstrap method has the property that the result depends on the choice of the operators in Eq.~\eqref{ops-sample} for constructing the bootstrap matrix. For example, we can use the following operators \cite{Berenstein:2021dyf}\footnote{In Eqs.~\eqref{ops-sample} and $\eqref{bootstrap}$, we considered the $K \times K $ bootstrap matrix. Here, we consider the $(K+1) \times (K+1)$ bootstrap matrix. In this way, $K$ denotes the size or the size minus one of the bootstrap matrix in this paper.}:
\begin{align}
	\tilde{O}_x :=  \sum_{n=0}^{K} c_{n} x^n.
	\label{operators-X}
\end{align}
Then, the bootstrap matrix becomes
\begin{align}
	{\mathcal M}_x^{(HO)}:=
	\begin{pmatrix}
		1 & \left\langle x \right\rangle  & \left\langle x^2 \right\rangle   & \cdots  \\
		\left\langle x	\right\rangle & \left\langle x^2	\right\rangle & \left\langle x^3 \right\rangle &  \cdots  \\
		\left\langle x^2	\right\rangle  &  \left\langle x^3 \right\rangle & \left\langle x^4 \right\rangle  & \cdots   \\
		\vdots & \vdots  & \vdots  & \ddots \\
	\end{pmatrix}
=
	\begin{pmatrix}
		1 & 0  &  E  & \cdots  \\
		0 &  E & 0 &  \cdots  \\
		E  &  0 & \frac{3}{2} E^2+ \frac{3}{8}   & \cdots   \\
		\vdots & \vdots  & \vdots  & \ddots \\
	\end{pmatrix}, 
	\label{bootstrap-X}
\end{align}
where we have used the relation \eqref{eq-recurrence-HO} in the second equality \cite{Aikawa:2021qbl}.
We can solve the condition ${\mathcal M}_x^{(HO)} \succeq 0$ by investigating Eq.~\eqref{eq-bootstrap-char}, and the result is shown in Table \ref{Table-HO-X}. We see that the results are similar to the anharmonic oscillator case, and the exact results cannot be derived. 

Therefore, the bootstrap method does not always reproduce the exact result. We have to choose an appropriate set of operators in the bootstrap matrix. We will discuss the difference between the operators $\tilde{O}_{xp}$ and $\tilde{O}_x$ in more detail in Sec.~\ref{sec-HO-bootstrap}.

\begin{table}
	\centering
	\begin{tabular}{|c||c|c|c|c|}
		\hline
		& $n= 0$   & $n= 1$  & $n= 2$  & $n= 3$ \\ 
		\hline
		\hline
		$K=1$ &   \multicolumn{4}{c |}{$0 \le E $}\\
		\hline
		$K=5$ &   \multicolumn{4}{c |}{$0.41745 \le E $}\\
		\hline
		$K=6$ &  $0.41745 \le E  \le 0.56651$ &   \multicolumn{3}{c |}{$1.1318 \le E $}    \\
		\hline 
		$K=9$ &  $0.49665 \le E  \le  0.50900 $ & $1.3677 \le E  \le 1.7781 $ &   \multicolumn{2}{c |}{$1.9625 \le E $}    \\
		\hline 
		$K=14$ &  $0.49992 \le E  \le  0.50001 $ & $1.4976 \le E  \le 1.5069 $ & $2.4284 \le E  \le 2.5656 $ &  $3.2861 \le E $    \\
		\hline 
		exact & 0.5 & 1.5 & 2.5 &  3.5   \\
		\hline 
	\end{tabular}
	\caption{Energy spectrum of the first four eigenstates ($n=0,1,2,3$) of the harmonic oscillator \eqref{H-HO} through the numerical bootstrap analysis.
		We take $\{ x^n \}$, ($n=0,1,\cdots,K $) in Eq.~\eqref{operators-X} for constructing the bootstrap matrix ${\mathcal M}_{x}$ \eqref{bootstrap-X}.
		As $K$ increases, the isolated allowed regions shrink, and tend to converge to the exact results.
		These properties are similar to the anharmonic oscillator case shown in Table \ref{Table-AHO-XP}.
	}
	\label{Table-HO-X}
\end{table}

\subsubsection{Other observables}
\label{sec-bootstrap-other-observables}

We have shown that the bootstrap method constraints the possible values of the observables $E$ and $\{\braket{x^n}\}_{n=1,\cdots,N_{\rm max}}$ for the energy eigenstates. Once we obtain these values, we can also obtain other polynomial type observables $\Braket{x^{i} p^{j} x^{k} p^l \cdots }$, where $i$, $j$, $k$, $l$ are non-negative integers by using the commutator relation $[x,p]=i$ and the recurrence relation \eqref{eq-xp-recurrence}.
For non-polynomial type observables, such as $\Braket{e^{ x}}$, if we construct the bootstrap matrix including these operators, it might be possible to obtain the allowed regions for these observables as well.
However, it is unclear when we obtain useful results through the bootstrap method for various observables in general, and we do not pursue this issue in this paper.
(Actually, we will see one difficulty in the hyperbolic Scarf potential in Appendix \ref{sec-HS}. There, although the energy eigenvalues can be obtained, the allowed regions of some observables such as $\Braket{1/\cosh x}$ do not converge to point like regions.)



\subsection{Properties of the allowed regions and their boundaries}
\label{sec-bootstrap-summary}

We have explored the basic idea of the bootstrap method and its applications to the harmonic and anharmonic oscillators. In these examples, we have obtained the energy eigenvalues from the allowed regions in which ${\mathcal M} \succeq 0$ is satisfied.
In order to conduct a more thorough analysis of the bootstrap method, it is helpful to summarize the terms that describe the properties of the allowed regions:

\begin{dfn}
{\it An allowed region} is the parameter region in which the bootstrap matrix is positive-semidefinite, i.e., ${\mathcal M} \succeq 0$.

\end{dfn}

\begin{dfn}
{\it An isolated allowed region} is the allowed region disconnected to other allowed regions.
It typically appears around the values corresponding to the energy eigenstates, such as the energy eigenvalue.
\end{dfn}

\begin{dfn}
{\it An allowed point} is a single point isolated allowed region in solvable systems.
\end{dfn}

Note that these regions depend on the size of the bootstrap matrix, and the allowed regions do not become larger as the size of the bootstrap matrix increases.
The proof is simple. Suppose that we construct a bootstrap matrix ${\mathcal M}$ from some operators $\{O_m\}$ ($m=1,\cdots, K$) as in Eq.~\eqref{bootstrap}. Now, by setting some of $c_n=0$ in Eq.~\eqref{ops-sample}, we can construct a smaller bootstrap matrix ${\mathcal M}'$. Obviously, the constraint ${\mathcal M}' \succeq 0$ is a subset of the constraint ${\mathcal M} \succeq 0$, and the allowed region of ${\mathcal M}$ is contained in the allowed region of ${\mathcal M}'$. Therefore, the allowed region does not become larger as the size of the bootstrap matrix increases.  \qed

Indeed we have seen that the allowed regions shrink in the anharmonic oscillator case.
On the other hand, in the harmonic oscillator case, once the allowed points are obtained, they are independent of the size of the bootstrap matrix  (for example, the point $E=1/2$ in Table \ref{Table-HO-XP} does not change as the size increases).
Thus, the existence of the allowed points independent of the size of the bootstrap matrix is a necessary condition to obtain (at least some of) the energy eigenvalues exactly by the bootstrap method.

It is therefore convenient to define the boundary of the allowed region as a ``rigid boundary", which has the following properties\footnote{The boundary of the allowed region in the bootstrap method is always ``rigid" in the sense that the expectation values cannot take values outside the boundary (although this boundary may shrink as the size $K$ increases). This rigidity and the rigidity of the rigid boundary defined here are used in different senses.}:

\begin{dfn}
{\it A rigid boundary} is a boundary (point) of an allowed region. Once it is obtained, it does not change as the size of the bootstrap matrix increases.
\end{dfn}

In the case of the harmonic oscillator, $E=n+1/2$, ($n=0,1,2,\cdots$) are rigid boundaries as we can see in Table \ref{Table-HO-XP}. 
In the case of the anharmonic oscillator, on the other hand, there are no rigid boundaries as shown in Table \ref{Table-AHO-XP}.
In this way, the rigid boundaries would generally appear in solvable systems.

Actually, by investigating the rigid boundary, we may obtain crucial properties of the energy eigenstate. 
In the next subsection, we will demonstrate that the information of the annihilation operators can be extracted from the rigid boundary in the harmonic oscillator case.
Thus, the bootstrap method tells us why the harmonic oscillator is solvable in terms of the creation-annihilation operators.

\subsubsection{Investigating rigid boundaries}
\label{sec-boundaries}

As we have discussed, rigid boundaries must exist in solvable systems. Thus, it would be important to investigate their properties. 

First, we argue when a boundary point is rigid. Suppose that we construct a $K \times K$ bootstrap matrix ${\mathcal M}$ from operators $\{O_m\}$ ($m=1,\cdots, K$) as in Eq.~\eqref{bootstrap}, i.e., ${\mathcal M}_{mn}=\langle O_m^\dagger O_n \rangle$.
At the boundaries of the allowed regions in which the bootstrap matrix is positive-semidefinite, at least one of the eigenvalues of the bootstrap matrix becomes zero.
Let us focus on a single point on the boundary and define the number of the zero eigenvalues of the bootstrap matrix at this point as $N_{\rm B}$.
We also define the corresponding $N_B$ eigenvectors as $\vec{v}^{(a)}$ ($a=1,\cdots, N_{\rm B}$), which satisfy ${\mathcal M}_{\rm B} \vec{v}^{(a)}=0$, where ${\mathcal M}_{\rm B}$ is the bootstrap matrix evaluated at this point.
This means that the following equations hold,
\begin{align} 
\sum_{m=1}^K	\langle O_k^\dagger O_m \rangle_{\rm B} v^{(a)}_m =0. \quad (a=1,\cdots, N_{\rm B},~\text{and}~ k=1,\cdots, K),
	\label{eq-boundary-eigen}
\end{align}
where $\langle O_k^\dagger O_m \rangle_{\rm B}$ is the value of $\langle O_k^\dagger O_m \rangle $ at the boundary point.
Here, we define the following operators for convenience,
\begin{align} 
\hat{V}^{(a)} := \sum_{m=1}^K O_m v^{(a)}_m , \quad (a=1,\cdots, N_{\rm B}).
	\label{eq-boundary-op}
\end{align}
Then, Eq.~\eqref{eq-boundary-eigen} can be written as
\begin{align} 
	\langle O_k^\dagger \hat{V}^{(a)} \rangle_{\rm B} =0.
	\label{eq-boundary-eigen2}
\end{align}

Suppose that we construct a $2 \times 2$ bootstrap matrix \eqref{bootstrap} from one of the operator $\{ \hat{V}^{(a)} \}$ and an arbitrary operator $O$. Then this bootstrap matrix at the boundary becomes
\begin{align}
	\begin{pmatrix}
		\langle \hat{V}^{(a)\dagger} \hat{V}^{(a)} \rangle_{\rm B}  & \langle  \hat{V}^{(a)\dagger} O \rangle_{\rm B}      \\
		\langle O^\dagger \hat{V}^{(a)}  \rangle_{\rm B}  &  \langle O^\dagger O \rangle_{\rm B}  
	\end{pmatrix}
	=	\begin{pmatrix}
		0  & \langle  \hat{V}^{(a)\dagger} O \rangle_{\rm B}      \\
		\langle O^\dagger \hat{V}^{(a)}  \rangle_{\rm B}  &  \langle O^\dagger O \rangle_{\rm B}  
	\end{pmatrix}, \quad (a=1,\cdots, N_{\rm B}),
	\label{bootstrap-boundary}
\end{align}
where we have used $\langle \hat{V}^{(a)\dagger} \hat{V}^{(a)} \rangle_{\rm B}=0$ obtained from Eq.~\eqref{eq-boundary-eigen2}.
Since the determinant of this bootstrap matrix is $-|\langle O^\dagger \hat{V}^{(a)}  \rangle_{\rm B}   |^2 \le 0 $, the eigenvalues are positive and negative and the matrix is not positive-semidefinite unless $\langle O^\dagger \hat{V}^{(a)}  \rangle_{\rm B} =0$.
Therefore, satisfying $\langle   O^\dagger \hat{V}^{(a)} \rangle_{\rm B} =0$ for ${}^\forall O$ is a necessary condition for the rigid boundary (otherwise the bootstrap matrix is not positive-semidefinite there and this boundary point is excluded from the original allowed region).

If the boundary point is rigid, the energy eigenstate corresponds to this boundary point should exist.
We denote this state as $\ket{{\rm B}}$. Then, the condition $\langle   O^\dagger \hat{V}^{(a)} \rangle_{\rm B} =0$ for ${}^\forall O$ becomes
\begin{align} 
	\hat{V}^{(a)} 	\ket{{\rm B}} =0, \quad (a=1,\cdots, N_{\rm B}).
	\label{eq-boundary-state}
\end{align}
Thus, all the operator $\hat{V}^{(a)}$ annihilate the eigenstate $\ket{{\rm B}}$, and they may characterize the rigid boundary\footnote{This discussion does not claim that the conventional annihilation operators such as $a=(x+ip)/\sqrt{2}$ in the harmonic oscillator are required to obtain rigid boundaries in the bootstrap method. The operator $\hat{V}^{(a)}$ might be of the form $\hat{V}^{(a)}= \hat{O}( H-E_B )$, where $E_B$ is the energy eigenvalue satisfying $ H \ket{{\rm B}} =E_B \ket{{\rm B}}  $  and $\hat{O}$ is an operator. Then Eq.~\eqref{eq-boundary-state} is trivially satisfied.}. (Note that if $N_{\rm B}  \ge 2$,  $\vec{v}^{(a)}$ degenerate and the operator $\hat{V}^{(a)}$ is not unique. Also, $N_{\rm B}$ and the form of $\hat{V}^{(a)}$ may depend on the size $K$.)\\

To better understand these properties of the rigid boundary and the operators $\hat{V}^{(a)}$, let us examine the harmonic oscillator.
We again consider the $5 \times 5$ bootstrap matrix \eqref{bootstrap-HO} constructed from the operators $\{O_m\}:=\{1,x,p,xp,x^2\}$.
As we have seen in Eq.~\eqref{eq-bootstrap-char-HO}, the bootstrap matrix \eqref{bootstrap-HO} is positive-semidefinite only if $E=1/2$ or $E \ge 3/2$.
Thus, the boundaries of these regions are $E=1/2$ and $E = 3/2$, and they are rigid boundaries as we have discussed. We will investigate the operators $\hat{V}^{(a)}$ at these two boundaries.

First we consider the point $E=1/2$.
The bootstrap matrix \eqref{bootstrap-HO} at $E=1/2$ becomes
\begin{align}
\left.	{\mathcal M} \right|_{E=1/2}=
	\begin{pmatrix}
		1 & 0  & 0   &  \frac{i}{2} & \frac{1}{2}  \\
		0 & \frac{1}{2} &  \frac{i}{2} & 0 &  0   \\
		0  & - \frac{i}{2} & \frac{1}{2}  & 0 & 0    \\
		-  \frac{i}{2}  &  0 & 0  &  \frac{3}{4} &   -\frac{3}{4} i   \\
		\frac{1}{2} & 0 & 0 &   \frac{3}{4} i&   \frac{3}{4}   \\
	\end{pmatrix}
	.
\end{align}
This matrix should have zero eigenvalues, and we indeed find two zero eigenvectors,
\begin{align}
	\vec{v}^{(1)}={}^t(0,1,i,0,0), \quad
	\vec{v}^{(2)}={}^t(0,0,0,i,1).
\end{align}
Then, we obtain the operators $\hat{V}^{(a)}$ defined in Eq.~\eqref{eq-boundary-op} as
\begin{align} 
  \hat{V}^{(1)}=O_m v_m^{(1)} =x+ip, \quad \hat{V}^{(2)}=O_m v_m^{(2)} =x(x+ip) =x   \hat{V}^{(1)}.
  \label{eq-HO-V-1/2}
\end{align}
Since $E=1/2$ is the rigid boundary, the energy eigenstate at $E=1/2$ (denoted as $\ket{{\rm B}}|_{E=1/2} $) is annihilated by these operators as in Eq.~\eqref{eq-boundary-state}:
\begin{align} 
	\hat{V}^{(a)}  \ket{{\rm B}}|_{E=1/2} =0, \quad (a=1,2).
\end{align}
From Eq.~\eqref{eq-HO-V-1/2}, this condition leads to $\hat{V}^{(1)}  \ket{{\rm B}}|_{E=1/2} =0$.
We see that $\hat{V}^{(1)}/\sqrt{2} = (x+ip)/\sqrt{2}:=a$ is the well known annihilation operator of the harmonic oscillator, and this condition reproduces the relation $a \ket{{\rm B}}|_{E=1/2} =0$ for the ground state in the harmonic oscillator.

Next, we consider the point $E=3/2$. The bootstrap matrix \eqref{bootstrap-HO} at $E=3/2$ becomes
\begin{align}
	\left.	{\mathcal M} \right|_{E=3/2}=
		\begin{pmatrix}
			1 & 0  & 0   &  \frac{i}{2} & \frac{3}{2}  \\
			0 & \frac{3}{2} &  \frac{i}{2} & 0 &  0   \\
			0  & - \frac{i}{2} & \frac{3}{2}  & 0 & 0    \\
			-  \frac{i}{2}  &  0 & 0  &  \frac{7}{4} &   -\frac{9}{4} i   \\
			\frac{3}{2} & 0 & 0 &   \frac{9}{4} i&   \frac{15}{4}   \\
		\end{pmatrix}
		.
	\end{align}
This matrix has one zero eigenvalue, and the eigenvector and the operator $\hat{V}^{(1)}$ are given by 
\begin{align}
	\vec{v}^{(1)}={}^t(-1,0,0,i,1), \quad \hat{V}^{(1)}=O_m v_m^{(1)} =-1+ixp+x^2=a^2+(H-3/2).
\end{align}
Then the energy eigenstate at $E=3/2$ should satisfy the condition
\begin{align} 
	\hat{V}^{(1)} \ket{{\rm B}}|_{E=3/2} = a^2  \ket{{\rm B}}|_{E=3/2}= 0,
\end{align}
where we have used $H \ket{{\rm B}} |_{E=3/2} = \frac{3}{2} \ket{{\rm B}} |_{E=3/2} $.
This condition reproduces the relation $a^2 \ket{E=3/2}=0$ for the first excited state in the harmonic oscillator.

Thus, by studying the rigid boundaries, we find that the energy eigenstates are annihilated by the annihilation operators $a^n$ ($n=1,2$). This means that the bootstrap method naturally leads to the creation-annihilation operator method, even if we do not know it from the beginning. Thus, investigating the rigid boundaries in the bootstrap method may provide important information about the system.

\section{Proposal: Detecting solvable systems through the numerical bootstrap method}
\label{sec-proposal}

As we have seen in the previous section, the results of the bootstrap method for solvable and unsolvable systems differ significantly. Using this property, there is a possibility that the numerical bootstrap method can be used as a tool for detecting solvable systems through the following steps:
\begin{enumerate}
	\item Prepare the (approximate) energy eigenvalues of the system, which we want to examine the solvability, by using any numerical method.
	
	\item Analyze the system using the bootstrap method\footnote{As we have argued in Sec.~\ref{sec-HO-AHO}, to obtain the allowed region, we have two options: the numerical linear programming (Sec.~\ref{sec-SDP}) and the characteristic polynomial method (Sec.~\ref{sec-SDP-char}). It may be efficient to use the linear programming method first. Although the numerical linear programming cannot provide the  strict allowed points in solvable systems, since they may be smeared due to numerical error, it is not difficult to distinguish the smeared allowed point in solvable systems from the isolated allowed region in non-solvable systems. We argue the details in Appendix \ref{app-exact-numerical}. Once we find a possible candidate for the allowed point, by applying the characteristic polynomial method, we may find the exact solution analytically.}. We start with a small size bootstrap matrix and gradually increase the size. 

	\item If we obtain the rigid boundaries (in particular the allowed points correspond to energy eigenstates), this means that the system is solvable. 
	
	Note that if any allowed region is not obtained even around the prepared approximate energy eigenvalue, it may signal that the system is solvable. This is the case where the bootstrap method misses the single point corresponding to the allowed point in the search space.

	\item If any allowed regions with rigid boundaries are not obtained, the system may not be solvable. However, if we change the bootstrap matrix constructed from a different set of operators, exact solutions might be obtained. (We will show in Sec.~\ref{sec-SI-bootstrap-energy} that choosing various operators is better.)
	
\end{enumerate}
Through this procedure, the bootstrap method can detect solvable systems. (However, we cannot conclude that the system is not solvable from the results of the bootstrap method alone. Obtaining exact results from the bootstrap method is a sufficient condition for the system to be solvable.)
In addition, the condition for the rigid boundary \eqref{eq-boundary-state} may tell us the properties of the energy eigenstates, such as the existence of annihilation operators, and we may understand the mechanism of the solvability of the system.\\

However, this is an optimistic proposal based on the systems for which the exact solutions have been obtained, such as the harmonic oscillator and P\"{o}schl-Teller potentials, and there is no guarantee that the exact solutions will always be obtained in other solvable models. There is a possibility that the allowed regions with rigid boundaries are not obtained as in the unsolvable cases even in solvable systems. To figure out this point, we study shape invariance, which is known as a sufficient condition for solvability and which many solvable system including harmonic oscillators and  P\"{o}schl-Teller potentials satisfy. We will show that if the system is shape invariant, the bootstrap method leads to the exact solutions if the bootstrap matrix is constructed from sufficiently various operators.

\section{Review of shape invariance}
\label{sec-SI}

We begin with a brief review of shape invariance \cite{Gendenshtein:1983skv}. For simplicity, we consider one-dimensional systems.

Shape invariance is known as a sufficient condition for a system to be solvable. 
A shape invariant system satisfies the following properties:
\begin{itemize}
	\item There are two sets of $N$ real parameters $\lambda:=(\lambda_1,\cdots,\lambda_N)$ and $\delta:=(\delta_1,\cdots,\delta_N)$.
	\item The Hamiltonian $\mathcal{H}$ can be expressed as
\begin{align} 
	\mathcal{H}=\mathcal{A}^{\dagger}(\lambda) \mathcal{A}(\lambda),
	\label{eq-SI-H}
\end{align}
where $\mathcal{A}(\lambda)$ is an annihilation operator depending on $\lambda$ and 
the ground state $\ket{0}_\lambda$ satisfies
\begin{align} 
  \mathcal{A}(\lambda)\ket{0}_\lambda=0.
  \label{eq-SI-0}
\end{align}
Note that we have set the ground state energy to be zero. (We distinguish the Hamiltonian $\mathcal{H}$, which has the zero ground state energy, from the usual Hamiltonian $H$.)
\item 
$\mathcal{A}(\lambda)$ satisfies the relation
\begin{align} 
	\mathcal{A(\lambda)}\mathcal{A}^{\dagger}(\lambda)=\mathcal{A}^{\dagger}(\lambda+\delta)\mathcal{A(\lambda+\delta)}+\epsilon(\lambda),
	\label{eq-SI}
\end{align}
where $\epsilon(\lambda)$ is a real function of $\lambda$ corresponding to the energy eigenvalue of the first excited state of the system, as we will soon show.
\end{itemize}

By using these properties, the energy eigenvalues and eigenstates of the system can be obtained as follows.
First, we introduce the following notation for a non-negative integer $n$:
\begin{align} 
	\mathcal{A}_n:=
	\mathcal{A}(\lambda+n\delta), \quad \epsilon_n := \epsilon(\lambda+n\delta), \quad \ket{0}_n:= \ket{0}_{\lambda+n\delta}.
\end{align}
Then, the Hamiltonian \eqref{eq-SI-H} is expressed as $\mathcal{H}=\mathcal{A}^{\dagger}_0 \mathcal{A}_0$, and the state $\ket{0}_n$ satisfies $\mathcal{A}_n \ket{0}_n=0$ through Eq.~\eqref{eq-SI-0}.
We introduce the following bracket $\{*,*\}$ for convenience:
\begin{align} 
	\{\mathcal{A}_n,\mathcal{A}_n^{\dagger}\}:=\mathcal{A}_n\mathcal{A}_n^{\dagger}-\mathcal{A}_{n+1}^{\dagger}\mathcal{A}_{n+1}=\epsilon_n.
	\label{eq-SI-2}
	\end{align}
In the second equality, we have used Eq.~\eqref{eq-SI} $n$ times.
Now, the Hamiltonian satisfies the commutation-like relation,
\begin{align} 
\mathcal{H}\mathcal{A}_0^{\dagger}\mathcal{A}_1^{\dagger}\cdots\mathcal{A}_{n-1}^{\dagger}=& \mathcal{A}_0^{\dagger} {\color{red} \mathcal{A}_0\mathcal{A}_0^{\dagger}}\mathcal{A}_1^{\dagger}\cdots\mathcal{A}_{n-1}^{\dagger} \nonumber \\
=& \mathcal{A}_0^{\dagger} {\color{red} \{\mathcal{A}_0,\mathcal{A}_0^{\dagger}\} }\mathcal{A}_1^{\dagger}\cdots\mathcal{A}_{n-1}^{\dagger}+\mathcal{A}_0^{\dagger} {\color{red} \mathcal{A}_1^{\dagger}\mathcal{A}_1}\mathcal{A}_1^{\dagger}\cdots\mathcal{A}_{n-1}^{\dagger} \nonumber \\
=& \mathcal{A}_0^{\dagger}\{\mathcal{A}_0,\mathcal{A}_0^{\dagger}\}\mathcal{A}_1^{\dagger}\cdots\mathcal{A}_{n-1}^{\dagger}+\mathcal{A}_0^{\dagger}\mathcal{A}_1^{\dagger}\{\mathcal{A}_1,\mathcal{A}_1^{\dagger}\}\mathcal{A}_2^{\dagger}\cdots\mathcal{A}_{n-1}^{\dagger} \nonumber \\
& + \cdots +
\mathcal{A}_0^{\dagger}\mathcal{A}_1^{\dagger}\cdots \mathcal{A}_{n-2}^{\dagger}
\{\mathcal{A}_{n-1},\mathcal{A}_{n-1}^{\dagger}\}+
\mathcal{A}_0^{\dagger}\mathcal{A}_1^{\dagger}\cdots\mathcal{A}_n^{\dagger}\mathcal{A}_n \nonumber \\
=& E_n \mathcal{A}_0^{\dagger}\mathcal{A}_1^{\dagger}\cdots\mathcal{A}_{n-1}^{\dagger}+\mathcal{A}_0^{\dagger}\mathcal{A}_1^{\dagger}\cdots\mathcal{A}_n^{\dagger}\mathcal{A}_n,
\label{eq-H-a} \\
&E_n:=\sum_{k=0}^{n-1}\epsilon_{k}.
\end{align}
Here we have used the relation \eqref{eq-SI-2} for the red colored terms in the second equality, and repeated it in the third equality.
From this relation and $\mathcal{A}_n \ket{0}_n=0$, the $n$-th excited state $\ket{n}$ and its energy eigenvalue $E_n$ of $\mathcal{H}$ are obtained as
\begin{align} 
	\ket{n} \propto	\mathcal{A}_0^{\dagger}\mathcal{A}_1^{\dagger}\cdots\mathcal{A}_{n-1}^{\dagger}\ket{0}_n, \quad E_n=\sum_{k=0}^{n-1}\epsilon_k.
	\label{eq-SI-E}
\end{align}
Taking $n=1$, $E_1=\epsilon_0=\epsilon(\lambda)$ is the energy eigenvalue of the first excited state, as we have mentioned.
We define $E_0:=0$ for later convenience.

Note that shape invariance does not guarantee that the obtained state $\ket{n}$ is physical. For example, if $E_n$ has a maximum at a certain level $n$, the states beyond that level are unphysical \cite{Sasaki:2014tka}.  This occurs when the number of the bound states is finite. We will see such unphysical states in Morse potentials in Sec.~\ref{sec-examples-bootstrap}. 

In this way, the energy eigenvalues and eigenstates of the shape invariant systems can be obtained analytically. However, it is generally difficult to determine whether a system has shape invariance. Also, not all solvable systems have shape invariance, and shape invariance is a sufficient condition for solvability. Thus, determining whether a system is solvable or not is generally difficult.

\section{Bootstrapping shape invariant systems}
\label{sec-SI-bootstrap}

\subsection{Derivation of the energy eigenvalues in shape invariant systems via bootstrap}
\label{sec-SI-bootstrap-energy}

In this section, we apply the bootstrap method to shape invariant systems. We will show that the bootstrap method reproduces the exact energy eigenvalues, if the bootstrap matrix is constructed from appropriate operators.

We consider a shape invariant system with the Hamiltonian
\begin{align} 
	\mathcal{H}=\mathcal{A}^{\dagger}_0 \mathcal{A}_0,
\end{align}
where $ \mathcal{A}_0$ satisfies the shape invariant condition \eqref{eq-SI-2}．

To use the bootstrap method, we need to choose the operators to construct the bootstrap matrix. In Sec.~\ref{sec-boundaries}, we have seen that the operators $\{ a^{n+1} \}$, which annihilate the eigenstates $|n \rangle$, are important to obtain the rigid boundaries in the harmonic oscillator case.
This suggests that the operators that annihilate the eigenstates $\ket{n}$ \eqref{eq-SI-E} in the shape invariant system are also important in the bootstrap method. We can easily show that the relation
\begin{align} 
	\mathcal{A}_{n} \mathcal{A}_{n-1} \cdots \mathcal{A}_1 \mathcal{A}_0 \ket{n} =0
	\label{eq-annihilation-n-SI}
\end{align}
is satisfied by using Eqs.\eqref{eq-SI-2} and \eqref{eq-SI-E}, and this motivates us to construct the bootstrap matrix from the operator,
\begin{align}
	\tilde{O}_\mathcal{A} := \sum_{n=0}^{K} c_{n} \left(  \mathcal{A}_{n-1}
	\mathcal{A}_{n-2} \cdots \mathcal{A}_1 \mathcal{A}_0 \right).
	\label{eq-O_A}
	\end{align}
	Here we have set $\mathcal{A}_{-1}:=I$, where $I$ is the identity operator. Then, the bootstrap matrix \eqref{bootstrap} is given by
\begin{align} 
	{\mathcal M}_\mathcal{A} =
	\begin{pmatrix}
	1 & \Braket{\mathcal{A}_0} & \Braket{\mathcal{A}_1\mathcal{A}_0} & \cdots \\
	\Braket{\mathcal{A}_0^{\dagger}} & \Braket{\mathcal{A}_0^{\dagger} \mathcal{A}_0} & \Braket{\mathcal{A}_0^{\dagger} \mathcal{A}_1\mathcal{A}_0} & \cdots \\
	\Braket{\mathcal{A}_0^{\dagger}\mathcal{A}_1^{\dagger}} & \Braket{\mathcal{A}_0^{\dagger}\mathcal{A}_1^{\dagger}\mathcal{A}_0} & \Braket{\mathcal{A}_0^{\dagger}\mathcal{A}_1^{\dagger}\mathcal{A}_1\mathcal{A}_0} & \cdots \\
	\vdots &  \vdots & \vdots & \ddots \\
\end{pmatrix},
\label{eq-SI-M}
\end{align}
and the $(m,n)$ component of this matrix is
\begin{align} 
{\mathcal M}_{mn}=\Braket{\mathcal{A}_0^{\dagger}\mathcal{A}_1^{\dagger}\cdots\mathcal{A}_{m-2}^{\dagger}\mathcal{A}_{n-2}\mathcal{A}_{n-3}\cdots\mathcal{A}_0}, \quad (1 \le n,m \le K+1).
\label{eq-SI-Mmn}
\end{align}

Now we investigate the allowed regions of the energy eigenvalue $E$ satisfying the condition ${\mathcal M}_\mathcal{A} \succeq 0$. A necessary condition for ${\mathcal M}_\mathcal{A} \succeq 0$ is that all the diagonal components ${\mathcal M}_{nn}$ are non-negative\footnote{We can show that ${\mathcal M}_{nn} \ge 0$ is a necessary condition for ${\mathcal M}_\mathcal{A} \succeq 0$ by taking all $c_m=0$ except $c_n$ in Eq.~\eqref{eq-O_A}.}.
Therefore, we focus on the diagonal components:
\begin{align} 
{\mathcal M}_{nn}=\Braket{ {\color{red} \mathcal{A}_0^{\dagger}\mathcal{A}_1^{\dagger}\cdots \mathcal{A}_{n-3}^{\dagger} \mathcal{A}_{n-2}^{\dagger}\mathcal{A}_{n-2} } \mathcal{A}_{n-3}\cdots \mathcal{A}_1 \mathcal{A}_0}.
\end{align}
Here, by applying the relation \eqref{eq-H-a}, the red colored terms in this equation becomes
\begin{align} 
\mathcal{A}_0^{\dagger}\mathcal{A}_1^{\dagger}\cdots\mathcal{A}_{n-3}^{\dagger}\mathcal{A}_{n-2}^{\dagger}\mathcal{A}_{n-2}=&
\left(\mathcal{H} -E_{n-2} \right)  \mathcal{A}_0^{\dagger}\mathcal{A}_1^{\dagger}\cdots\mathcal{A}_{n-3}^{\dagger}.
\label{eq-SI-A-HE}
\end{align}
Since we are considering the energy eigenstate with the energy eigenvalue $E$, $\langle \mathcal{H} \cdots \rangle = \langle E \cdots \rangle$ is satisfied, and it leads to 
\begin{align} 
	{\mathcal M}_{nn}= (E-E_{n-2}){\mathcal M}_{n-1,n-1}.
\end{align}
By repeating this relation, we obtain
\begin{align} 
	{\mathcal M}_{nn} =  (E-E_{n-2})(E-E_{n-3})\cdots(E-E_1)E,
	\label{eq-SI-Mnn}
\end{align}
where we have used ${\mathcal M}_{22}= \Braket{\mathcal{A}_0^{\dagger} \mathcal{A}_0} = \Braket{\mathcal{H}} = E$.
Thus, the condition that all ${\mathcal M}_{nn}$ are non-negative becomes
\begin{align} 
 (E-E_{n-2})(E-E_{n-3})\cdots(E-E_1)E \ge 0, \quad n=2,3,\cdots,K+1.
\label{eq-SI-Mnn2}
\end{align}
If $E_{n+1}>E_n$ for ${}^\forall n \le K-2$, this condition requires
\begin{align} 
	E=0=E_0,~E=E_1,~\cdots,~E=E_{K-2}~\text{or}~E\geq E_{K-1}.
	\label{eq-SI-E-bootstrap}
	\end{align}
This means that the allowed values of the energy eigenvalue $E$ are limited to $E_n$, and it is consistent with the result of shape invariance \eqref{eq-SI-E}.
In particular, by taking $K \to \infty$, all $E_n$ are reproduced.
These are, of course, the allowed points with respect to $E$.

If the condition $E_{n+1}>E_n$ is not satisfied, the inequalities \eqref{eq-SI-Mnn2} tell us that $E=E_{n+1}$ is not allowed, unless $E_{n+1}$ coincides with an $E_m$ $(m \le n)$. Thus, $E_{n+1}$ does not correspond to a new energy eigenvalue. Since the inequality $E_{n+1} \le E_n$ means that there is a local maximum at a certain $m <n+1$, the state $\ket{n+1}$ is unphysical according to the discussion at the end of Sec.~\ref{sec-SI}. Therefore, the bootstrap method is consistent with this discussion. We will study this problem further in the Morse potential case in Sec.~\ref{sec-Morse}. \\

We have seen that the bootstrap method exactly reproduces the energy eigenvalues of the shape invariant system if we use the bootstrap matrix ${\mathcal M}_\mathcal{A}$ \eqref{eq-SI-M} constructed from the annihilation operators $\mathcal{A}_n$ \eqref{eq-annihilation-n-SI}. However, in practice, if we knew the annihilation operators $\mathcal{A}_n$ from the beginning, we do not need to use the bootstrap method to obtain the energy eigenvalues (we can derive the spectrum through the conventional method argued in Sec.~\ref{sec-SI}). 
Therefore, a crucial question is whether the bootstrap method leads to the exact energy eigenvalues without using the annihilation operators $\mathcal{A}_n$.
The answer is yes, if we construct the bootstrap matrix using sufficiently various and many operators.

Suppose that we prepare a set of various well defined operators $\{O_n\}$ to construct the bootstrap matrix \eqref{bootstrap}.
If the operators are sufficiently many, the annihilation operators $ \{ \mathcal{A}_{n-1} \mathcal{A}_{n-2} \cdots \mathcal{A}_1 \mathcal{A}_0  \}$ \eqref{eq-annihilation-n-SI} would be expressed as a linear combination of the operators $\{O_n\}$.
Then, the operator $\tilde{O}$ in Eq.~\eqref{ops-sample} would be written as
\begin{align}
	\tilde{O}=  \sum_{n=1}^{K} c_{n} O_n=  \left( \sum_{n=1}^{\tilde{K}} \tilde{c}_{n} \mathcal{A}_{n-2} \mathcal{A}_{n-3} \cdots \mathcal{A}_1 \mathcal{A}_0 \right) +\left(\sum_{n=\tilde{K}+1}^{K} \tilde{c}_{n} \tilde{O}_n  \right) ,
\label{eq-O-A-SI}
\end{align}
where $\tilde{K}$ is an integer less than or equal to $K$, $\{ \tilde{c}_{n} \}$ are constants mapped from $\{ c_{n} \}$ by a bijective linear map, and $ \{ \tilde{O}_n \}$ are certain operators. Thus, the condition that the $K \times K$ bootstrap matrix constructed from $\{O_n\}$ is positive-semidefinite is stronger than that of the $\tilde{K} \times \tilde{K}$ bootstrap matrix constructed from $\{\mathcal{A}_{n-1} \mathcal{A}_{n-2} \cdots \mathcal{A}_1 \mathcal{A}_0\}$. Since the latter bootstrap matrix provides the exact energy eigenvalues, the former does as well. Here, importantly, the bootstrap method by using the operators $\{O_n\}$ automatically leads to the exact energy eigenvalues, even if we do not know the concrete form of the right-hand side of Eq.~\eqref{eq-O-A-SI}. (Recall that  $\{ c_{n} \}$ are arbitrary constants and we do not have to specify  $\{ \tilde{c}_{n} \}$.) Therefore, we do not need to know the annihilation operators $\mathcal{A}_n$ from the beginning in the bootstrap method.\\

We have shown that the bootstrap method leads to the exact energy eigenvalues of the shape invariant systems, if we employ sufficiently various and many operators $\{O_n\}$ to construct the bootstrap matrix. This is the main conclusion of this work. 

\subsection{Rigid boundaries and obtaining the annihilation operators}
\label{sec-exact-boundary-SI}

In Sec.~\ref{sec-boundaries}, we have seen that the rigid boundary at $E=E_n$ in the harmonic oscillator case is related to the annihilation operator $a^{n+1}$.
We can find similar relation in the shape invariant systems.

The diagonal component of the bootstrap matrix ${\mathcal M}_\mathcal{A} $ satisfies Eq.~\eqref{eq-SI-Mnn},
\begin{align} 
	{\mathcal M}_{nn}=&\Braket{\mathcal{A}_0^{\dagger}\mathcal{A}_1^{\dagger}\cdots \mathcal{A}_{n-3}^{\dagger} \mathcal{A}_{n-2}^{\dagger}\mathcal{A}_{n-2}  \mathcal{A}_{n-3}\cdots \mathcal{A}_1 \mathcal{A}_0} \nonumber \\
	= & (E-E_{n-2})(E-E_{n-3})\cdots(E-E_1)E, \quad (n=2,3,\cdots,K+1).
\end{align}
Thus, at $E=E_n$, we obtain
\begin{align} 
	\begin{cases}
		\mathcal{M}_{mm} |_{E=E_n}\neq 0 &(m=1,2,\cdots, n+1 ),\\
		\mathcal{M}_{mm} |_{E=E_n}= 0&(m = n+2, n+3,\cdots,K+1).\\
	\end{cases} 
\end{align}
Here $\mathcal{M}_{mm} |_{E=E_n} = 0$ would correspond to the zero eigenvalue of the bootstrap matrix.
Since $E=E_n$ is an rigid boundary (assuming that $E=E_n$ is physical), this relation implies that the energy eigenstate at $E=E_n$ satisfies
\begin{align} 
	\mathcal{A}_{n} \cdots \mathcal{A}_1 \mathcal{A}_0 \ket{E_n}=0,
	\label{eq-exact-boundary-SI}
\end{align}
as in Eq.~\eqref{eq-boundary-state}.
This equation
 reproduces Eq.~\eqref{eq-annihilation-n-SI}. 
Thus, the annihilation operators can be obtained from the rigid boundaries in the bootstrap method generally in shape invariant systems.

As we have studied in the previous subsection, the bootstrap method does not require the specific expressions of the annihilation operators from the beginning to obtain the energy spectrum. Eq.~\eqref{eq-exact-boundary-SI} implies that the information of the annihilation operators is encoded in the bootstrap matrix at the rigid boundaries, and we can read off the annihilation operators from the zero eigenvectors $\vec{v}^{(a)}$ \eqref{eq-boundary-eigen}. This explains the results of the harmonic oscillator in Sec.~\ref{sec-boundaries}. This mechanism would be useful when we explore unknown solvable systems.

\section{Examples of bootstrap analyses of shape invariant systems}
\label{sec-examples-bootstrap}

In this section we investigate harmonic oscillators and Morse potentials by using the bootstrap method as concrete examples of shape invariant systems. In the case of the Morse potentials, the energy eigenstates \eqref{eq-SI-E} derived from shape invariance include unphysical states related to the violation of the condition $E_{n+1}>E_n$. We will show that such states are automatically removed in the bootstrap method.
We also show the bootstrap analyses of Rosen-Morse potentials and hyperbolic Scarf potentials, which are also shape invariant systems in Appendix \ref{app-SI-bootstrap}.

\subsection{Bootstrapping harmonic oscillators again}

In the case of harmonic oscillators, as we have already seen in Sec.~\ref{sec-HO-AHO}, the exact results are derived by the numerical bootstrap method using the bootstrap matrix ${\mathcal M}_{xp}$ \eqref{bootstrap-XP} constructed from the operators $\{x^mp^n\}$ \eqref{operators-XP}. However, it is unclear why the exact results are obtained there. On the other hand, we have also seen that shape invariance is a key to deriving the exact results. Therefore, it will be valuable to study the harmonic oscillator in terms of shape invariance. Note that the bootstrap analysis in this subsection is a development of the previous work \cite{Aikawa:2021qbl}. 

\subsubsection{Shape invariance in the harmonic oscillator}

The Hamiltonian of the harmonic oscillator \eqref{H-HO} is expressed as
\begin{align} 
	\mathcal{H}= a^\dagger a \left(=H-\frac{1}{2}\right),
\end{align}
by using the annihilation operator $a:=(x+ip)/\sqrt{2} $.
Here this operator satisfies the relation
\begin{align} 
aa^{\dagger}=a^{\dagger}a+1.
\end{align}
By comparing this relation with the shape invariant condition \eqref{eq-SI}, we find that there are no parameters corresponding to $\lambda$ and $\delta$ in the harmonic oscillator, and we obtain $\mathcal{A}_n=a$ and $\epsilon_n=1$. Thus, the general formula \eqref{eq-SI-E} for the energy eigenstates and eigenvalues in shape invariance leads to 
\begin{align} 
	\ket{n} \propto (a^{\dagger})^n \ket{0}, \quad E_n=n ,
	\label{E-HO}
\end{align}
which is the well known relation in quantum mechanics.

\subsubsection{Bootstrapping the harmonic oscillator with the annihilation operator $a$}
\label{sec-HO-bootstrap}

We apply the bootstrap method to the harmonic oscillator by using shape invariance discussed in Sec.~\ref{sec-SI-bootstrap}. We employ the bootstrap matrix \eqref{eq-SI-M} constructed from the annihilation operator $a$,  
\begin{align} 
\mathcal{M}_{\mathcal{A}}^{(HO)}=
\begin{pmatrix}
	\Braket{1} & \Braket{a} & \Braket{a^2} & \cdots \\
\Braket{a^{\dagger}} & \Braket{a^{\dagger}a} & \Braket{a^{\dagger}a^2} & \cdots \\
\Braket{(a^{\dagger})^2} & \Braket{(a^{\dagger})^2 a} & \Braket{(a^{\dagger})^2 a^2} & \cdots \\
\vdots &  \vdots & \vdots & \ddots \\
\end{pmatrix} .
	\label{bootstrap-HO-A}
\end{align}
The diagonal components of this matrix are calculated as 
\begin{align} 
\mathcal{M}_{nn}=
\begin{cases}
1&(n=1),\\
E(E-1)(E-2)\cdots(E-(n-2))\quad&(n\geq2),
\end{cases} 
\end{align}
by using Eq.~\eqref{eq-SI-Mnn} and $E_n=n$.
Since all the diagonal components has to be non-negative, it is possible only when $E=0,1,2,\cdots$.
Thus, the bootstrap method reproduces the correct results \eqref{E-HO}, and we confirm that the general formula \eqref{eq-SI-E-bootstrap} works explicitly.\\

Although we have already obtained the energy eigenvalues, we evaluate the off-diagonal components of the bootstrap matrix $\mathcal{M}_{\mathcal{A}}^{(HO)}$ \eqref{bootstrap-HO-A} to understand the properties of the whole bootstrap matrix.
By substituting $O=(a^{\dagger})^m a^n$ into the equation \eqref{HO=0}, we obtain
\begin{align}
	&\Braket{\left[a^{\dagger}a,(a^{\dagger})^m a^n\right]}=0 \quad
	\Rightarrow \quad (m-n)\Braket{(a^{\dagger})^m a^n}=0.
	\end{align}
This relation leads to
	\begin{align} 
		\mathcal{M}_{m+1,n+1}=	\Braket{(a^{\dagger})^m a^n}=0, \quad (m\neq n).
	\end{align}
	Thus, the off-diagonal components of $\mathcal{M}_{\mathcal{A}}^{(HO)}$ are simply all zero, and it becomes
	\begin{align} 
		\mathcal{M}_{\mathcal{A}}^{(HO)}=
		\begin{pmatrix}
			1 &  \\
	 & E & & \text{\Huge O} \\
	 & & E(E-1) &  \\
	 \text{\Huge O} & & & \ddots \\
	 \end{pmatrix}.
	\label{bootstrap-HO-A2}
	 \end{align}
Therefore, the off-diagonal components of the bootstrap matrix does not provide any conditions on the spectrum.\\

We have shown analytically that the bootstrap method reproduces the exact results of the harmonic oscillator by using the bootstrap matrix $\mathcal{M}_{\mathcal{A}}^{(HO)}$ \eqref{bootstrap-HO-A} constructed from the annihilation operators $\{a^n \}$. 
As we have argued in Sec.~\ref{sec-SI-bootstrap-energy}, even if we do not use the annihilation operators $\{a^n \}$, using the operators $\{O_m\}$ is sufficient, if the operators $\{a^n \}$ can be expressed as a linear combination of $\{O_m\}$ as in Eq.~\eqref{eq-O-A-SI}.
In fact, we have studied in Sec.~\ref{sec-HO-AHO} that the bootstrap analysis using the operators $\{x^mp^n\}$ leads to the exact results.
Obviously, the operators $\{x^mp^n\}$ satisfy the above condition, and this explains why we obtain the exact results there.

Let us confirm it more explicitly by focusing on the case of $(K_x,K_p)=(1,1)$ in Table \ref{Table-HO-XP}, where we have used the operators $O_m=\{1,x,p,xp\}$ in Eq.~\eqref{operators-XP} and obtained the constraint $E \ge 1/2$.
By using these operators, Eq.~\eqref{eq-O-A-SI} can be written as
\begin{align}
	\tilde{O}= c_1 1+ c_2x+c_3p+c_4 xp = &  c_1 1+c_2 (x+ip) + (c_3-ic_2) p +c_4 xp \nonumber \\
	=&\left( \sum_{n=1}^2 \tilde{c}_n a^{n-1} \right)+ \tilde{c}_3 p +\tilde{c}_4 xp  ,
	\label{eq-xp-A-HO}
\end{align}
where $\{\tilde{c}_n\}:=\{c_1, \sqrt{2}c_2,c_3-ic_2,c_4\}$ is related to $\{c_n\}$ by a bijective linear map.
Thus, the constraint of the bootstrap method at $(K_x,K_p)=(1,1)$ is stronger than that obtained from the $2\times 2$ version of the matrix $\mathcal{M}_{\mathcal{A}}^{(HO)}$ \eqref{bootstrap-HO-A2} constructed from $\{1,a\}$.
In fact, the constraint $\mathcal{M}_{\mathcal{A}}^{(HO)} \succeq 0$ leads to $E \ge 0$, and this is equivalent to $E\ge 1/2$ at $(K_x,K_p)=(1,1)$ (we have shifted $E \to E-1/2$ in this section such that $E_0=0$ ). 
Therefore, as we expected, the results of the bootstrap method using the operators $\{x^mp^n\}$ are related to those of the annihilation operators $\{a^n\}$.

On the other hand, we have also seen in Sec.~\ref{sec-HO-caution} that the bootstrap analysis using the operators $\{x^n\}$ does not reproduce the exact results. We presume that this is because the operators $\{x^n\}$ is not sufficient to express the annihilation operators $\{a^n \}$ as in Eq.~\eqref{eq-xp-A-HO}.


\subsection{Bootstrapping Morse potential}
\label{sec-Morse}

\subsubsection{Review of Morse potential and shape invariance}
\label{sec-Morse-SI}
\begin{figure}[t]
	\centering
	\includegraphics[scale=0.7]{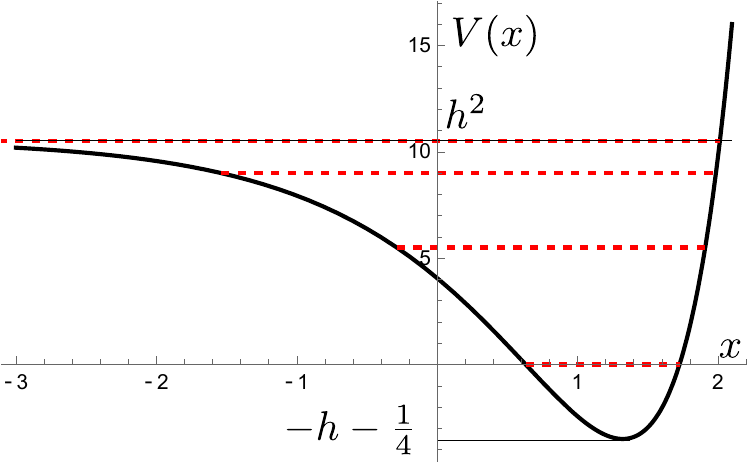}
	\caption{The Morse potential \eqref{eq-Morse-H} at $(h,\mu)=(13/4,1)$. The red dashed lines represent the energy eigenvalues \eqref{eq-Morse-E2} of the bound states (four bound states appear at $h=13/4=3.25$). The bottom of the potential is set to be $-h-1/4$ (at $\mu e^x=h+1/2$) such that the ground state energy $E_0$ is zero. The continuous spectrum appears in $E\geq h^2$.}
\label{fig-Morse-pot}
\end{figure}

We consider the Morse potential \cite{PhysRev.34.57, Landau:1991wop}, where the Hamiltonian is given by
	\begin{align} 
		\mathcal{H}=p^2+V(x), \quad V(x)=\mu^2 e^{2x}-\mu(2h+1)e^x+h^2.
	\label{eq-Morse-H}  
	\end{align}
Here we have taken the kinetic term $p^2$ instead of $p^2/2$ for simplicity. $\mu$ and $h$ are real parameters satisfying $\mu >0$ and $h >0$\footnote{If $h\le 0$, no bound state appears in this model.}. However, $\mu$ can be eliminated by the translation of $x$, and is not an important parameter physically. The potential is depicted in Fig.~\ref{fig-Morse-pot}. Since $V(x) \to h^2$ as $x\to -\infty$, there is no bound state in $E\geq h^2$ and the continuous spectrum appears there. We have fixed the constant term of the potential $V(x) $ such that the ground state energy is zero as we will see soon.

We show shape invariance of this system. First, we introduce the annihilation operator $\mathcal{A}_0$ as
\begin{align} 
\mathcal{A}_0=ip+\mu e^x-h,\quad \mathcal{A}_0^{\dagger}=-ip+\mu e^x-h,
\end{align}
and express the Hamiltonian \eqref{eq-Morse-H} as
\begin{align} 
	\mathcal{H}=&\mathcal{A}_0^{\dagger}\mathcal{A}_0   .
	\label{eq-Morse-H-A}
\end{align}
Here the annihilation operator $\mathcal{A}_0$ satisfies
\begin{align} 
	\mathcal{A}_0 \mathcal{A}_0^{\dagger}=&p^2+\mu^2 e^{2x}-\mu(2h-1)e^x+h^2 \nonumber \\
=& \left(-ip+\mu e^x-h+1\right)\left(ip+\mu e^x-h+1\right) +2h-1 .
\end{align}
Thus, we define
\begin{align} 
	\mathcal{A}_1:=ip+\mu e^x-(h-1),\quad \mathcal{A}_1^{\dagger}:=-ip+\mu e^x-(h-1), \quad \epsilon_1=2h-1,
\end{align}
and they satisfy the shape invariant condition \eqref{eq-SI} with $\lambda=h$, $\delta=-1$, and $\epsilon(\lambda)=2h-1$.
Therefore, we obtain
\begin{align} 
	\mathcal{A}_n:=ip+\mu e^x-(h-n),\quad \mathcal{A}_n^{\dagger}:=-ip+\mu e^x-(h-n), \quad \epsilon_n=2(h-n)+1.
	\label{eq-Morse-An}
\end{align}
By using these operators, the energy eigenstates $\ket{n}$ and eigenvalues $E_n$ are obtained through the general formula \eqref{eq-SI-E}. In particular, the energy eigenvalues are given by
\begin{align} 
	E_n=\sum_{k=0}^{n-1}\left(2(h-k)-1\right)=h^2-(h-n)^2.
	\label{eq-Morse-E}
\end{align}
Note that, if we regard $n$ as a continuous parameter,  the energy eigenvalue $E_n$ increases for $n<h$, while it decreases for $n>h$. 
Thus, depending on the value of $h$, $E_{[ h]' +1}$ is less than $E_{[ h]'}$, where $[ h]'$ is the greatest integer less than $h$ (max$\{ n \in {\mathbf Z}| n<h  \}$), e.g. $[ 3.01]'=3$ and $[ 3]'=2$.
In fact, it is known that the bound states are up to $n=[ h]'$ and the energy eigenstates beyond $E=E_{[ h]'}$ are the continuum of states in the region $E \ge h^2$.
Thus, the energy eigenvalues are summarized as
\begin{align} 
	E_0=0,~E_1,~\cdots,~E_{[ h]'},\text{ and }E \ge h^2.
	\label{eq-Morse-E2}
\end{align}
Although the general formula \eqref{eq-SI-E} predicts the energy eigenstates $\ket{n}$ even for $n>[ h]'$, they are unphysical (they satisfy the Schr\"{o}dinger equation but are non-normalizable \cite{Odake:2013vma}) and do not appear in the spectrum \eqref{eq-Morse-E2}.

We can also compute $\Braket{e^x}$ for each eigenstates as \cite{https://doi.org/10.1002/qua.560310205}
\begin{align}
	\begin{cases}
	\Braket{ e^x}|_{E=E_n} =\frac{h-n}{\mu} , \quad &n=0,1,\cdots,[ h]', \\
	\Braket{ e^x}|_{E} =0  , \quad &E \ge h^2.
	\end{cases}
	\label{eq-Morse-X}
\end{align}
Although no bound state appears in the region $E \ge h^2$, the expectation value $\Braket{e^x} $ for the non-normalizable mode is zero, because the wave function of this mode spreads over the region $x \to -\infty$, where $e^x \to 0$\footnote{\label{ftnt-non-normalizable}A more precise explanation is as follows: For $E \ge h^2$, the energy eigenstate $\Ket{E}$ does not belong to the Hilbert space because its norm  $\langle E | E \rangle$  diverges, and the expectation value $\Braket{ e^x}|_{E}$ does not make sense. Therefore, to evaluate $\Braket{ e^x}|_{E}$, we must regularize $| E \rangle$ such that $\langle E | E \rangle$ is a real, positive, finite quantity. Then, we can compute the expectation value as $\Bra{E} e^x \Ket{E}/\langle E | E \rangle$. In this case, it is clear that for $E \ge h^2$, $\Braket{ e^x}|_{E}=0$.}.

In the next subsection, we will show that the bootstrap method reproduces the energy eigenvalues \eqref{eq-Morse-E2} and $\Braket{e^x}$ \eqref{eq-Morse-X}. 

\subsubsection{Bootstrap analysis of the Mose potential. }

We apply the bootstrap method to investigate the Morse potential \eqref{eq-Morse-H}. 
We employ the following operators
\begin{align}
	\tilde{O}:=  \sum_{m=0}^{K_x} \sum_{n=0}^{K_p} c_{mn} e^{mx} p^n
	\label{operators-XP-Morse}
\end{align}
in Eq.~\eqref{ops-sample} to construct the bootstrap matrix and examine the bootstrap method. Note that, since the Morse potential is expressed by the exponential function $e^{x}$ and $e^{2x}$, $\{ e^{mx} \}$ is more natural operator than $\{ x^m \}$, which we used in the polynomial potential cases \eqref{operators-XP}. 
Then, the bootstrap matrix \eqref{bootstrap} is given by
\begin{align} 
{\mathcal M}^{(\text{Morse})}=
\begin{pmatrix}
\Braket{1} & \Braket{e^x} & \Braket{p}  & \cdots \\
\Braket{e^x} & \Braket{e^{2x}} & \Braket{e^x p} & \cdots \\
\Braket{p} & \Braket{pe^x} & \Braket{p^2} & \cdots \\
\vdots &  \vdots & \vdots  & \ddots \\
\end{pmatrix}
=
\begin{pmatrix}
1 & \Braket{e^x} &  0 & \cdots \\
\Braket{e^x} & \frac{2h+1}{2\mu} \Braket{e^x}  & \frac{i}{2}\Braket{e^x}  & \cdots \\
 0 & -\frac{i}{2} \Braket{e^x} & E - h^2 + \frac{2h+1}{2} \mu\Braket{e^x}  & \cdots \\
\vdots &  \vdots & \vdots  & \ddots \\
\end{pmatrix} .
\label{eq-bootstrap-matrix-Morse-XP}
\end{align}
Here, we have solved the constraints \eqref{HO=0} and \eqref{HO=EO} by computer in the second equation (See Appendix \ref{app-bootstrap-details} for the details of the derivation).
Similar to the bootstrap matrix of the anharmonic oscillator ${\mathcal M}_{xp}$ in Eq.~\eqref{bootstrap-XP2}, all the components of ${\mathcal M}^{(\text{Morse})}$ are expressed by $E$, $\Braket{e^x}$ and $\Braket{p}$ in the Morse potential case, but we have set $\Braket{p}=0$ for simplicity, which is satisfied in this potential\footnote{Obviously, there are no degenerate states in this potential, and the wave function can be real and $\Braket{p}=0$ is satisfied. Besides, we can perform the bootstrap method without assuming $\Braket{p}=0$ and obtain the consistent results.}.

\begin{figure}[t]
	\centering
	\includegraphics[scale=0.7]{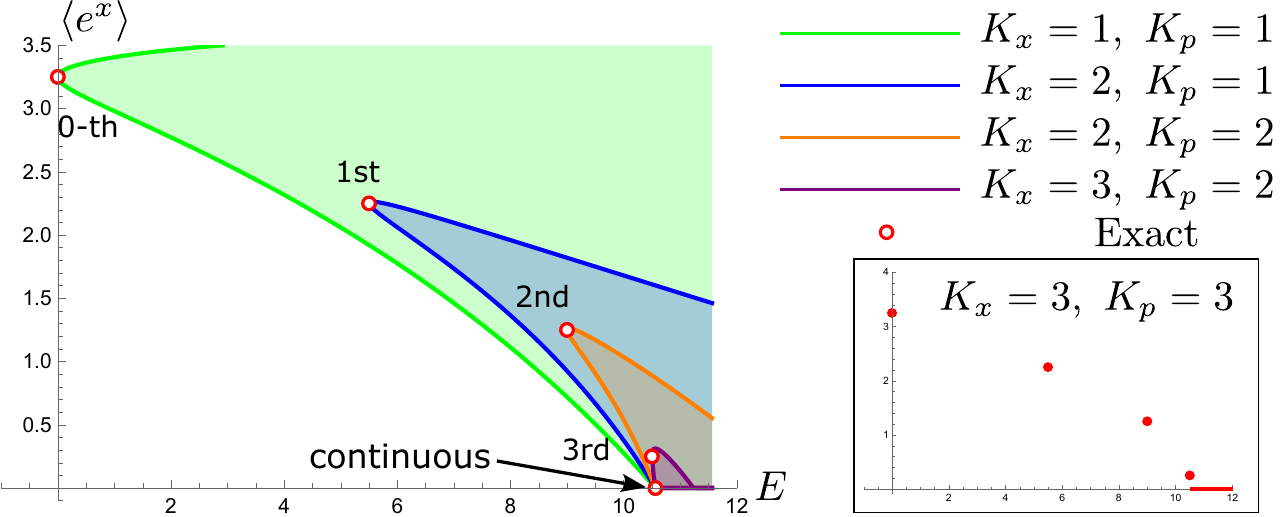}
	\caption{Bootstrap analysis of the Morse potential \eqref{eq-Morse-H} at $(h,\mu)=(13/4,1)$. We use the operators $\{ e^{mx} p^n \}$ ($m=0,1,\cdots,K_x $ and $n=0,1,\cdots,K_p $) in Eq.~\eqref{operators-XP-Morse} to construct the bootstrap matrix ${\mathcal M}^{(\text{Morse})}$ \eqref{eq-bootstrap-matrix-Morse-XP}, and solve the linear programming discussed in Sec.~\ref{sec-SDP}. At $h=13/4=3.25$, $[h]' =3$ and four bound states appear.
 The small red circles represent the exact energy eigenstates \eqref{eq-Morse-E2} (the continuous spectrum $E>h^2$ is omitted). The inside of the curves and the dots represent the allowed regions where the energy eigenstates can exist. 
	At $(K_x,K_p)=(3,3)$, the numerical bootstrap method reproduces the known analytic results including the continuous spectrum. Since some allowed points are not visible because they overlap, see Table \ref{Table-Morse-XP} for the details. Note that the allowed points in the numerical linear programming are slightly smeared due to numerical error. See Appendix \ref{app-exact-numerical} for more details.
	}
	\label{fig-Morse-XP}
\end{figure}

Using the linear programming analysis discussed in Sec.~\ref{sec-SDP}, we can numerically determine the allowed region of the bootstrap matrix ${\mathcal M}^{(\text{Morse})} \succeq 0$. The result at $(h,\mu)=(13/4,1)$ is shown in Fig.~\ref{fig-Morse-XP}. Due to numerical error, the allowed points are not exactly points but are smeared in the numerical linear programming (See the details in Appendix \ref{app-exact-numerical}). To obtain the exact allowed points, we solve the equation \eqref{eq-bootstrap-char} and \eqref{eq-bootstrap-char-deriv} obtained from the characteristic polynomial \eqref{eq-char-poly} by using the \texttt{Solve} function in Mathematica. Note that the \texttt{Solve} function can derive analytic solutions of algebraic equations, and the result is summarized in Table \ref{Table-Morse-XP}\footnote{We assume that the smeared allowed points obtained by the numerical linear programming, as shown in Fig.~\ref{fig-Morse-XP}, converge to the allowed points obtained by the characteristic polynomial, as shown in Table \ref{Table-Morse-XP}. However, we cannot completely exclude the possibility that this is not true.}.
The known analytic result \eqref{eq-Morse-E2} including the continuous spectrum $E \ge h^2$ is reproduced at $(K_x,K_p)=(3,3)$.
The expectation value $\Braket{e^x}$ \eqref{eq-Morse-X} is also obtained correctly. We have examined other values of $(h,\mu)$, and obtained the exact results, although we do not show them in this paper.\\

As we have argued in Sec.~\ref{sec-SI-bootstrap-energy}, the bootstrap method can reproduce the energy spectrum in shape invariant systems if $E_{n+1}>E_n$ is satisfied and operators satisfying the relation \eqref{eq-O-A-SI} are employed. Our operators \eqref{operators-XP-Morse} satisfy this condition, and obtaining the spectrum $E_n$ \eqref{eq-Morse-E} of the bound states is ensured.
However, we have also seen that the bootstrap method correctly reproduces the expectation value $\Braket{e^x}$ and the continuous spectrum $E \ge h^2$, which is not related to $E_n$ in Eq.~\eqref{eq-Morse-E}. These cannot be explained by the argument in Sec.~\ref{sec-SI-bootstrap-energy} alone.
We show how the bootstrap method derives these quantities in Appendix \ref{app-Morse}.

\begin{table}[t]
	\centering
	\begin{tabular}{|c||c|c|c|c|c|}
		\hline
		& $n= 0 $  & $n= 1 $ & $n= 2$  & $n= 3$ & continuous \\ 
		\hline
		\hline
		$(K_x,K_p)=(1,1)$ &   \multicolumn{5}{c |}{$0 \le E $}\\
		\hline 
		$(K_x,K_p)=(2,1)$ &  0 & \multicolumn{4}{c |}{$11/2 \le E $}\\
		\hline 
		$(K_x,K_p)=(2,2)$ &  0 &  11/2 & \multicolumn{3}{c |}{$9 \le E $}\\
		\hline 
		$(K_x,K_p)=(3,2)$ &  0 &  11/2 & 9 & \multicolumn{2}{c |}{$21/2 \le E $}\\
		\hline 
		$(K_x,K_p)=(3,3)$ &  0 &  11/2 & 9 &  21/2  & $169/16 \le E$ \\
		\hline 
		\hline 
		exact result for $E_n$~\eqref{eq-Morse-E}   & 0 & 11/2 & 9 &  21/2 &  $ 169/16 \le E$   \\
		\hline 
		$\Braket{e^x}|_{E=E_n}$~\eqref{eq-Morse-X} & 13/4 & 9/4 & 5/4 &  1/4 &  0   \\
		\hline 
	\end{tabular}
	\caption{Energy spectrum of the Morse potential \eqref{eq-Morse-H} at $(h,\mu)=(13/4,1)$ via the bootstrap method. We evaluate the characteristic polynomial of the bootstrap matrix ${\mathcal M}^{(\text{Morse})}$ discussed in Sec.~\ref{sec-SDP-char} by using Mathematica and obtain the allowed points, which rigorously agree with the energy spectrum \eqref{eq-Morse-E} and the values $\Braket{e^x}$ \eqref{eq-Morse-X}. 
	We obtain consistent results from the numerical linear programming shown in Fig. \ref{fig-Morse-XP}, but the allowed points are slightly smeared due to numerical error. 
	}
	\label{Table-Morse-XP}
\end{table}

\section{Discussions}
\label{sec-discussion}

In this study, we have shown that the bootstrap method can derive the exact solutions for systems with shape invariance. 
Therefore, even if the solvability of a given system is not known, we can examine whether the system is solvable by using the bootstrap method.
In particular, to determine the system is solvable, it would be sufficient to obtain low energy eigenvalues, and the size of the bootstrap matrix does not need to be very large. Thus, the numerical cost would be not so high.

Our formulation is independent of the details of the Hamiltonian. Thus, it will work even in unusual systems, e.g. solvable `discrete' quantum mechanics in which the momentum operator $p$ appears in the Hamiltonian as $e^{\gamma p}$ ($\gamma \in \mathbf{R}$) \cite{Odake:2008bf}. 

A future direction is to investigate whether the bootstrap method can derive the exact solutions for solvable systems that do not have shape invariance. In Appendix \ref{app-DT}, as a simple example, we consider the systems that are mapped from shape invariant systems by the Krein-Adler transformation \cite{zbMATH03242479, zbMATH00916231, Odake:2011jj, Odake:2012bc}.
Such systems are generally not shape invariant but are still solvable.
We show that the bootstrap method can derive the exact solutions of these systems.
Besides, it would be interesting to apply the bootstrap method to quasi exactly solvable systems, in which a finite number of energy eigenvalues can be explicitly obtained, while the rest cannot.
\\

The numerical bootstrap method may also derive the exact results in quantum many-body systems. 
(The application of the bootstrap method to many-body systems and field theories has been developed in Refs.\cite{Anderson:2016rcw, Lin:2020mme, Han:2020bkb, Morita:2022zuy, Bhattacharya:2021btd, Kazakov:2021lel, Lawrence:2021msm, Kazakov:2022xuh, Cho:2022lcj, Lin:2023owt, Kazakov:2024ool, Li:2024wrd, Berenstein:2024ebf, Li:2024ggr, Cho:2024kxn, Lin:2024vvg,Cho:2024owx, Guo:2025fii}.) For example, the ground state energy of free theories whose Hamiltonian are expressed by the creation-annihilation operators, would be reproduced exactly by using the bootstrap method \cite{Aikawa:2021qbl}, although these are trivial examples\footnote{See, for example, Refs.~\cite{10.21468/SciPostPhys.13.4.090, 10.21468/SciPostPhys.16.2.041} for other approaches to numerical dectection of integrable systems in quantum many body systems.}.
One interesting application is the gauge theories that have the gravity duals in the gauge/gravity correspondence \cite{Maldacena:1997re, Aharony:1999ti}. Some quantities in these systems are claimed to be ``solvable" by using gravity, and in fact the solvability of some of them has been shown on the gauge theory side as well. However, we do not know how to solve the many other quantities in the gauge theories. Therefore, it is interesting to apply the bootstrap method to these systems to clarify whether they are indeed solvable. The rigid boundaries may also tell us why they are solvable.

\paragraph*{Acknowledgment.---}
We would like to thank S.~Odake for valuable discussions and comments.
We would also like to thank an anonymous referee for her/his helpful comments and for providing us useful equations for the Rosen-Morse potential and the hyperbolic Scarf potential as mentioned in footnote \ref{ftnt-referee}.
The work of T.~M. is supported in part by Grant-in-Aid for Scientific Research C (No. 20K03946) from JSPS.

\appendix

\section{Some useful equations for the bootstrap analysis}
\label{app-bootstrap-details}

We have presented the concrete expressions of the bootstrap matrices for the several models in Eqs.~\eqref{bootstrap-XP2}, \eqref{bootstrap-HO}, \eqref{bootstrap-X} and \eqref{eq-bootstrap-matrix-Morse-XP}. To obtain these equations, we have to derive the relation \eqref{eq-pxp-x} by solving the recurrence relations \eqref{eq-xp-recurrence}. 
The derivation is straightforward but complicated, and we typically use computer to solve the recurrence relation.

This appendix introduces some useful equations for these computations in the harmonic oscillator, anharmonic oscillator, and Morse potential cases. We omit the Rosen-Morse and hyperbolic Scarf potential cases because the equations are too complicated.

In the cases of the harmonic and anharmonic oscillators, the components of the bootstrap matrix are given in a form $\langle p^l x^m p^n \rangle$. 
Through the commutation relation $[p,x]=-i$, this operator can be rewritten as
\begin{align} 
   p^l x^m p^n= \sum_{k=0}^{\min (l,m)  } (-i)^k  \binom{l}{k} \frac{m!}{(m-k)! } x^{m-k} p^{l+n-k} 
  \label{eq-xp-ordering}
 \end{align}
where  $\binom{l}{k}:= l!/k!(l-k)!$ is the binomial coefficient.
Then, every component of the bootstrap matrix can be expressed by the $\langle x^m p^n \rangle$ type expectation values, and these can be obtained by solving the recurrence relations \eqref{eq-xp-recurrence}.
By using the explicit form of $V(x)$, these recurrence equations are reduced to the following formulas for the harmonic oscillator 
\begin{align}
\langle x^m p^n \rangle &= 0, \quad  \text{if } m+n \text{ is odd}, \nonumber \\
\langle x^{m+1} \rangle &= \frac{1}{m+1}\left( 2mE \langle x^{m-1} \rangle + \frac{1}{4} m(m-1)(m-2) \langle x^{m-3} \rangle \right), \nonumber \\
\langle x^m p \rangle &= \frac{1}{2} i m \langle x^{m-1} \rangle, \nonumber \\
\langle x^m p^{n+2} \rangle &=  m(m-1) \langle x^{m-2} p^{n} \rangle + 2i m \langle x^{m-1} p^{n+1} \rangle +
 2E \langle x^m p^n \rangle - \langle x^{m+2} p^n \rangle ,
\label{eq-recurrence-HO}
\end{align}
and for the anharmonic oscillator,
\begin{align} 
\langle x^{m+3} \rangle &= 
\frac{1}{m+2}
\left( -2(m+1) \langle x^{m+1} \rangle 
+4 mE \langle x^{m-1} \rangle + \frac{1}{2} m(m-1)(m-2) \langle x^{m-3} \rangle
\right)
, \nonumber \\
\langle x^m p \rangle &= \frac{1}{2} i m \langle x^{m-1} \rangle, \nonumber \\
\langle x^m p^{n+2} \rangle &= m(m-1) \langle x^{m-2} p^{n} \rangle + 2i m \langle x^{m-1} p^{n+1} \rangle + 2E \langle x^m p^n \rangle - \langle x^{m+2} p^n \rangle - \frac{1}{2} \langle x^{m+4} p^n \rangle .
\label{eq-recurrence-AHO}
 \end{align}
In these equations, $m\langle x^{m-1} \rangle$, $m(m-1)(m-2)\langle x^{m-3} \rangle$, $n\langle x^{m+1}p^{n-1} \rangle$, and $n(n-1)\langle x^m p^{n-2} \rangle$ are zero when $m=0$, $m=0,1,2$, $n=0$, and $n=0,1$, respectively (i.e., $\langle x^{-1} \rangle$, $\langle x^{-2} \rangle$ are not defined).
These equations imply that $\langle x^m p^n \rangle$ are polynomials of $E$ in the harmonic oscillator case and are polynomials of $E$, $\langle x \rangle$ and $\langle x^2 \rangle$ in the anharmonic oscillator case.

Through a similar procedure, we obtain the following relation for the Morse potential \eqref{eq-Morse-H} case
\begin{align}
   p^l e^{mx} p^n &= \sum_{k=0}^{ l  } (-im)^k  \binom{l}{k}   e^{mx} p^{l+n-k}, \nonumber \\
\langle e^{(m+2)x} \rangle &= \frac{1}{2(m+1)\mu^2} \left[ (2m+1)\mu(2h+1)\langle e^{(m+1)x} \rangle + \frac{m}{2} \left( m^2 + 4(E - h^2) \right) \langle e^{mx} \rangle \right], \nonumber \\
\langle e^{mx}p \rangle &= \frac{i m}{2} \langle e^{mx} \rangle,~(m \neq 0), \nonumber \\
\langle e^{mx}p^{n+2} \rangle &= (E - h^2+m^2) \langle e^{mx}p^n \rangle
+ 2im \langle e^{mx}p^{n+1} \rangle + \mu (2h+1)  \langle e^{(m+1)x}p^n \rangle
 -\mu^2 \langle e^{(m+2)x}p^n \rangle.
\label{eq-recurrence-Morse}
\end{align}
These equations implies that $\langle e^{mx} p^n \rangle$ are polynomials of $E$, $\langle e^{x} \rangle$ and $\langle p \rangle$ in the Morse potential case.

\section{Details of our numerical linear programming}
\label{app-numerical}

\subsection{A sample Mathematica code for the linear programming of the anharmonic oscillator}
\label{app-sample-code}

 In this appendix, we provide a sample Mathematica code\footnote{The code can be found at the following URL: \url{https://www2.yukawa.kyoto-u.ac.jp/~takeshi.morita/}} for the bootstrap analysis of the anharmonic oscillator \eqref{H-AHO}. The code is designed to analyze the first $3\times3$ part of the bootstrap matrix ${\mathcal M}_{xp}^{(AHO)}$ of the anharmonic oscillator \eqref{bootstrap-XP2}, but it can be easily modified to analyze other potentials.

We define a symbol \texttt{xp[n,m]}:=$\langle x^n p^m \rangle$ and find the minimum and maximum of $\langle x^2 \rangle$= \texttt{xp[2,0]} for a given energy $E$. The code is as follows:

\begin{verbatim}
	(* Define the bootstrap matrix. *)
M = {{1, xp[1,0], 0}, {xp[1,0], xp[2,0], I/2}, {0, -I/2, 1/3*(4*ene - xp[2,0])}}

	(* Define the energy eigenvalue search space. *)
ene1 = 0;
ene2 = 1;
Nene = 100;
energylist = Subdivide[ene1, ene2, Nene];

	(* Choose the variable that you want to minimize. *)

minimizedvariable = xp[2, 0]

	(* Choose the solver and the tolerance. *)

method = {"DSDP", Tolerance -> 10^(-6)}

	(* The main code for the bootstrap analysis. *)

dataMin = {};
dataMax = {};
Do[
 bootstrapMatrix = M /. ene -> i;
 bootstrapvariables = Variables[bootstrapMatrix];
 
 constraints = 
  VectorGreaterEqual[{bootstrapMatrix, 0}, {"SemidefiniteCone", 
    Length[bootstrapMatrix[[1]]]}];

 tmp = SemidefiniteOptimization[minimizedvariable, {constraints}, 
   bootstrapvariables, Method -> method];

 If[NumberQ[tmp[[1, 2]]], AppendTo[dataMin, {i, tmp}]];

 tmp = SemidefiniteOptimization[-minimizedvariable, {constraints}, 
   bootstrapvariables, Method -> method];

 If[NumberQ[tmp[[1, 2]]], AppendTo[dataMax, {i, tmp}]];

 , {i, energylist}]

(* Plot the allowed region: energy vs. x^2. *)

ListPlot[{Table[{i[[1]], minimizedvariable /. i[[2]]}, {i, dataMax}], 
  Table[{i[[1]], minimizedvariable /. i[[2]]}, {i, dataMin}]}]

\end{verbatim}
The minimum and maximum values of $\langle x^2 \rangle$ at each energy are stored in the lists \texttt{dataMin} and \texttt{dataMax}, respectively. 
In this code, we use the function \texttt{SemidefiniteOptimization} to solve the semidefinite programming problem. This function is available versions in 12.0 and later. We used the solver \texttt{DSDP} in this code, but we use \texttt{MOSEK}, which must be installed separately, in the numerical analysis of this paper.

One important parameter in this code is a \texttt{Tolerance}, which is set to $10^{-6}$ in this sample.
To determine whether a bootstrap matrix is positive-semidefinite, there must be no negative eigenvalues. In numerical analysis, however, sufficiently small negative eigenvalues are approximated as zero. The accuracy of this approximation is controlled by the \texttt{Tolerance}: The smaller the \texttt{Tolerance}, the stricter the result. Thus, roughly speaking, a smaller \texttt{Tolerance} leads to a smaller allowed region.

In Appendix \ref{app-exact-numerical}, we will discuss that the \texttt{Tolerance} dependence of the results is crucial in determining whether the system is solvable.

In order to perform the bootstrap analysis for a larger bootstrap matrix \eqref{bootstrap-XP}, we need to add codes that compute the matrix element $\langle p^l x^m p^n \rangle$ by solving Eqs.~\eqref{eq-xp-ordering} and \eqref{eq-recurrence-AHO}.

\subsection{Smeared allowed points in numerical linear programming}
\label{app-exact-numerical}

This paper uses two methods for determining the allowed region: numerical linear programming (Sec.~\ref{sec-SDP}) and the characteristic polynomial analysis (Sec.~\ref{sec-SDP-char}).
The numerical programming is more convenient because it enables faster analysis and can be applied to larger bootstrap matrices and more independent variables. However, it cannot precisely determine the allowed points of solvable systems. In this appendix, we present the numerical results of solvable systems and demonstrate that the allowed points are indeed smeared. Nevertheless, we show that these smeared allowed points differ significantly from the isolated allowed points of non-solvable systems and are sufficient for distinguishing between them.

In order to distinguish these two isolated allowed regions, it is convenient to introduce the following definitions.
\begin{dfn}
{\it A smeared allowed point} is a numerically obtained isolated allowed region.
It is an allowed point in a solvable system, but it is smeared due to numerical error.
\end{dfn}

\begin{figure}[t]
	\centering
	\includegraphics[scale=0.5]{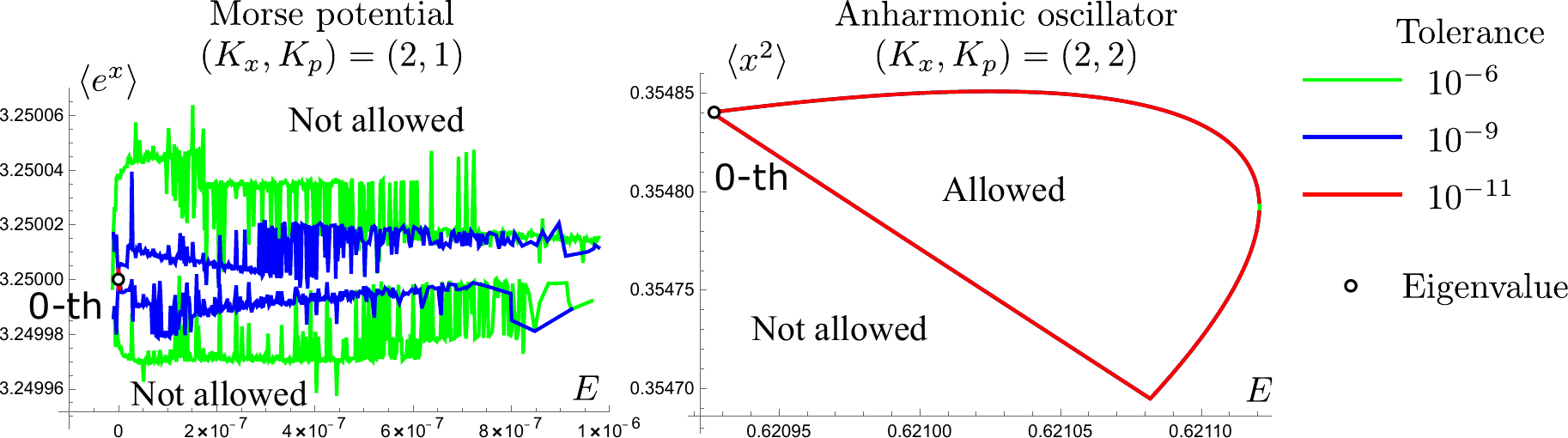}
	\caption{The \texttt{Tolerance} dependence of the allowed regions of the Morse potential and the anharmonic oscillator obtained by using the numerical linear programming. These are the enlarged view of Figs.~\ref{fig-x4} and \ref{fig-Morse-XP}. The curves represent the boundaries of the isolated allowed regions obtained by changing the \texttt{Tolerance} with a fixed size of the bootstrap matrix. In the anharmonic oscillator case, the curve is almost insensitive to the \texttt{Tolerance}, and the difference is almost invisible in the figure. On the other hand, in the Morse potential case, it strongly depends on the \texttt{Tolerance}, and converges to the exact result (the details are summarized in Table. \ref{Table-Morse-Tol}). Also, the boundary is jagged and noisy due to numerical error. Thus, the isolated allowed regions of the Morse potential and the anharmonic oscillator are significantly different.
	}
	\label{fig-Tol}
\end{figure}

In Fig.~\ref{fig-Tol}, we show the isolated allowed regions of the Morse potential and anharmonic oscillator obtained by using the numerical linear programming. These correspond to the ground states shown in Figs.~\ref{fig-x4} and \ref{fig-Morse-XP}.
We plot these isolated allowed regions for various values of the \texttt{Tolerance} parameter, which controls the numerical accuracy of the linear programming and is introduced in Appendix \ref{app-sample-code}. 

These figures show two notable differences between the Morse potential and the anharmonic oscillator:
\begin{enumerate}
	\item The isolated allowed region of the anharmonic oscillator is almost independent of the \texttt{Tolerance} parameter, while that of the Morse potential depends significantly on it. As we decrease \texttt{Tolerance}, the isolated allowed region of the Morse potential shrinks.
	The \texttt{Tolerance} dependence of the Morse potential is summarized in Table \ref{Table-Morse-Tol}.
		\item The boundary of the allowed region of the anharmonic oscillator is smooth, whereas the boundary of the allowed region of the Morse potential is jagged and noisy.
\end{enumerate}
These differences suggest that the isolated allowed region of the Morse potential is actually a single point, which is smeared by numerical error\footnote{The isolated allowed region of the unsolvable system also converges to an extremely narrow region, if the size of the bootstrap matrix is sufficiently large. However, analyzing large bootstrap matrices is more difficult numerically, and the boundary of the isolated allowed region may become noisy and the \texttt{Tolerance} dependence may be significant, similar to smeared allowed points in solvable systems. For solvable systems, however, the situation is quite different. Very small isolated allowed regions with the \texttt{Tolerance} dependence suddenly appear even for small bootstrap matrices.}.

In addition to these two differences, the allowed point in the solvable system is a rigid boundary that does not change when the size of the bootstrap matrix changes. This differs significantly from the isolated allowed region in the non-solvable system. Thus, although we cannot obtain the exact allowed point through the numerical linear programming, we can distinguish it from the isolated allowed region in the non-solvable system.

\begin{table}[t]
	\centering
	\begin{tabular}{|c||c|c|}
		\hline
Tolerance		&   energy at the ground state & $\braket{e^{x}}$  \\ 
		\hline
		\hline
		$10^{-6}$  &   $-1.32 \times 10^{-8} \le E \le 1.35 \times 10^{-6} $  &  $3.249956 \le \braket{e^{x}} \le 3.250065$  \\
		$10^{-9}$  &   $-9.18 \times 10^{-9} \le E \le 1.17 \times 10^{-6} $  &  $3.249978 \le \braket{e^{x}} \le 3.250026$  \\
		$10^{-11}$  &   $-5.38 \times 10^{-11} \le E \le 1.66 \times 10^{-9} $  &  $3.249997 \le \braket{e^{x}} \le 3.250003$  \\
		\hline 
\hline
		 Exact  &   $E=0 $  &  $ \braket{e^{x}}=13/4 = 3.25 $  \\
		\hline 
	\end{tabular}
	\caption{The \texttt{Tolerance} dependence of the numerical linear programming for the Morse potential \eqref{eq-Morse-H} at $(h,\mu)=(13/4,1)$. The maximum and minimum values of $E$ and $\braket{e^{x}}$ of the ground state shown in Fig.~\ref{fig-Tol} are summarized.
	As the \texttt{Tolerance} decreases, the allowed region shrinks and converges to the exact result.
	}
	\label{Table-Morse-Tol}
\end{table}

\section{Details of the bootstrap analysis of the Morse potential}
\label{app-Morse}

In Sec.~\ref{sec-SI-bootstrap}, we have shown that the bootstrap method can derive the exact spectrum of the shape invariant systems, if the condition $E_{n+1} > E_n $ is satisfied. 
However, this condition could be violated in the Morse potential \eqref{eq-Morse-H} when $n \ge [h]'$, and the continuous spectrum appears $E \ge h^2 $, where $[ h]'$ is the greatest integer less than $h$ as defined below Eq.~\eqref{eq-Morse-E}.
Nevertheless, the bootstrap method can reproduce the complete spectrum including the continuous spectrum.
Additionally, the values of $\langle e^x \rangle |_{E=E_n}$ \eqref{eq-Morse-X} are also reproduced as shown in Sec.~\ref{sec-Morse}.
This appendix explains how the bootstrap method derives these quantities.

\subsection{Derivation of the complete spectrum of the Morse potential}
\label{app-Morse-off-diagonal}

To show the derivation of the complete spectrum \eqref{eq-Morse-E2} through the bootstrap method, it is enough to consider the bootstrap matrix $\mathcal{M}_{\mathcal{A}}$ \eqref{eq-SI-M} constructed from the annihilation operators $\mathcal{A}_n$ of the Morse potential \eqref{eq-Morse-An} because of the relation \eqref{eq-O-A-SI}.
We construct the $(K+1)\times (K+1)$ bootstrap matrix $\mathcal{M}_{\mathcal{A}}$ \eqref{eq-SI-M} by using the annihilation operators $\mathcal{A}_n$ \eqref{eq-Morse-An}. 
Since all the diagonal components ${\mathcal M}_{nn}$ of $\mathcal{M}_{\mathcal{A}}$ have to be non-negative, we obtain the condition \eqref{eq-SI-Mnn2} for the energy eigenvalue $E$,
\begin{align} 
	{\mathcal M}_{nn}= (E-E_{n-2})(E-E_{n-3})\cdots(E-E_1)E \ge 0, \quad n=2,3,\cdots,K+1,
	\label{eq-Morse-Mnn-cond}
\end{align}
where $E_n$ is given by \eqref{eq-Morse-E}.
As we have studied in Sec.~\ref{sec-SI-bootstrap-energy}, these inequalities lead to the energy eigenvalues $E=E_n$, ($n \le K-2 $) \eqref{eq-SI-E-bootstrap}, when $E_{n}>E_{n-1}$ $({}^\forall n \le K-2)$ is satisfied. However, in the Morse potential case, $E_n$ is monotonically decreasing for $n >h$, and this condition is violated. Therefore, we must be careful in our analysis.

First we take $K=[h]' +1$. Then, since $E_n$ is monotonically increases at least up to $n= [h]'$, the assumption  $E_{n+1}>E_n$ is held, and the inequalities \eqref{eq-Morse-Mnn-cond} leads to the condition for $E$,
\begin{align} 
	E=0=E_0,~E=E_1,~\cdots,~E=E_{[h]'-1},\text{ or }E \ge E_{[h]'}.
	\label{eq-Morse-E-bootstrap}
	\end{align}

	Next we consider $K>[h]' +1$. In this case, the condition for $E$ from the inequalities \eqref{eq-Morse-Mnn-cond} is different depending on whether $E_{[h]'+1}$ is larger or smaller than $E_{[h]'}$, and we obtain\footnote{
The relation between $E_{[h]'+1}$ and $E_{[h]'}$ is determined by the value of $\tilde{h}$ that is the fractional part of $h$ ($h=[h]' + \tilde{h}$). If $\tilde{h}>1/2$, $\epsilon_{[h]'}$ defined in Eq.~\eqref{eq-Morse-An} is positive and $E_{[h]'+1} > E_{[h]'}$.}
	\begin{align} 
		\begin{cases}
		E=0,~E=E_1,~\cdots,~E=E_{[h]'-1},~E=E_{[h]'},\text{ or } E \ge E_{[h]'+1}, \quad & (E_{[h]'+1} > E_{[h]'})\\
		E=0,~E=E_1,~\cdots,~E=E_{[h]'-1},\text{ or }E \ge E_{[h]'}, & (E_{[h]'+1} \le E_{[h]'}).
		\end{cases}
			\label{eq-Morse-E-bootstrap-cond}
	\end{align}
This condition does not depend on $K$, if $K>[h]' +1$. This is because $E_{n+1}<E_n$ for ${}^\forall n>[h]' +1$, and $E_{n+1}$ takes a value in the region already prohibited by the condition \eqref{eq-Morse-E-bootstrap}.
Therefore, this is the final result obtained from the condition \eqref{eq-Morse-Mnn-cond} that all the diagonal components ${\mathcal M}_{nn}$ are non-negative. 

However, the actual energy eigenvalues \eqref{eq-Morse-E2} implies that we should obtain the condition 
\begin{align} 
	E=0,~E=E_1,~\cdots,~E=E_{[h]'-1},~E=E_{[h]'},\text{ or }E \ge h^2.
		\label{eq-Morse-E-bootstrap-final}
\end{align}
By comparing this condition with the obtained condition \eqref{eq-Morse-E-bootstrap-cond}, we find that they coincide except that the region $E_{[h]'} < E < h^2$ or $E_{[h]'+1} < E < h^2$ is allowed in Eq.~\eqref{eq-Morse-E-bootstrap-cond}. So these regions are redundant.

In fact, we can remove these redundant regions by evaluating the contribution of the off-diagonal components of the bootstrap matrix $\mathcal{M}_{\mathcal{A}}$ \eqref{eq-SI-M}. 
To show this, we do not need to consider the whole bootstrap matrix $\mathcal{M}_\mathcal{A} $, but considering only the following $2 \times 2$ submatrix $\tilde{\mathcal M}$ is sufficient,
\begin{align} 
	\tilde{\mathcal M}:=&
	\begin{pmatrix}
		{\mathcal M}_{n+1,n+1}  & {\mathcal M}_{n+1,n+2}  \\
		{\mathcal M}_{n+2,n+1} & {\mathcal M}_{n+2,n+2} 
	\end{pmatrix}	, \qquad (0 \le n \le K-1)	\nonumber \\
=&	\begin{pmatrix}
		\mathcal{A}_0^{\dagger}\mathcal{A}_1^{\dagger}\cdots\mathcal{A}_{n-1}^{\dagger}\mathcal{A}_{n-1}\mathcal{A}_{n-2}\cdots\mathcal{A}_{0} & \mathcal{A}_0^{\dagger}\mathcal{A}_1^{\dagger}\cdots\mathcal{A}_{n-1}^{\dagger}\mathcal{A}_{n}\mathcal{A}_{n-1}\cdots\mathcal{A}_{0} \\
	\mathcal{A}_0^{\dagger}\mathcal{A}_1^{\dagger}\cdots\mathcal{A}_{n}^{\dagger}\mathcal{A}_{n-1}\mathcal{A}_{n-2}\cdots\mathcal{A}_{0} & \mathcal{A}_0^{\dagger}\mathcal{A}_1^{\dagger}\cdots\mathcal{A}_{n}^{\dagger}\mathcal{A}_{n}\mathcal{A}_{n-1}\cdots\mathcal{A}_{0} 
	\label{eq-Morse-M-sub}
	\end{pmatrix}.
\end{align}
A necessary condition for $\mathcal{M}_\mathcal{A} \succeq 0 $ is that all the submatrices $\tilde{\mathcal M}$ $(n=0,\cdots,K-1)$ are positive semidefinite.

Let us investigate the condition that one submatrix $\tilde{\mathcal M}$ is positive semidefinite.
The diagonal components of $\tilde{\mathcal M}$ are the same as those of $\mathcal{M}_\mathcal{A}$ computed in Eq.~\eqref{eq-SI-Mnn}, and are given by
\begin{align}
	\tilde{\mathcal M}_{11}=&E(E-E_1)(E-E_2)\cdots(E-E_{n-1}) , \nonumber \\
	\tilde{\mathcal M}_{22}=&E(E-E_1)(E-E_2)\cdots(E-E_{n-1}) (E-E_{n}).
	\label{eq-Morse-M11}
\end{align}
The off-diagonal component $\tilde{\mathcal M}_{12}={\mathcal M}_{n+1,n+2}$ can be evaluated as follows.
Since the annihilation operator of the Morse potential is $\mathcal{A}_n= ip +\mu e^x-(h-n)$, it satisfies
\begin{align} 
[ \mathcal{A}_n,\mathcal{A}_{m}]=0.
\label{eq-Morse-A-n}
\end{align}
Thus, we obtain
\begin{align}
	{\mathcal M}_{n+1,n+2}
=&
\Braket{ \mathcal{A}_0^{\dagger}\mathcal{A}_1^{\dagger}\cdots\mathcal{A}_{n-1}^{\dagger}\mathcal{A}_{n-1} \mathcal{A}_{n-2}\cdots\mathcal{A}_{0}\mathcal{A}_{n}} \nonumber \\
=& E(E-E_1)(E-E_2)\cdots(E-E_{n-1}) \Braket{ \mathcal{A}_{n}},
\end{align}
where we have shifted $\mathcal{A}_{n}$ to the right and used Eq.\eqref{eq-SI-A-HE} repeatedly.
Then, $\tilde{\mathcal M}$ can be written as\footnote{\label{ftnt-general-SI}The equation \eqref{eq-sub-M} is valid for any systems with shape invariance that satisfy the following two conditions: (i) $ \mathcal{A}_n+ \mathcal{A}_n^\dagger=a_n+b_n f(x,p)$, where $a_n$ and $b_n$ are constants and $f(x,p)$ is a function of $x$ and $p$, and (ii) $\mathcal{A}_n- \mathcal{A}_n^\dagger=\mathcal{A}_{n-1}- \mathcal{A}_{n-1}^\dagger$. Then, it is not difficult to show that the recurrence relations of ${\mathcal M}_{n+1,n+2} \pm {\mathcal M}_{n+2, n+1}$ lead to Eq.~\eqref{eq-sub-M}.}
\begin{align} 
	\tilde{\mathcal M}=&E(E-E_1)\cdots(E-E_{n-1}) M, \quad 
	M:=
\begin{pmatrix}
	1 & \Braket{\mathcal{A}_n} \\
\Braket{\mathcal{A}_n^{\dagger}} & E-E_n
\end{pmatrix}.
\label{eq-sub-M}
\end{align}

From now on, we consider the situation that $E$ satisfies $E(E-E_1)\cdots(E-E_{n-1}) > 0$. 
Then, for $\tilde{\mathcal M}$ to be positive semidefinite, $M$ must be positive semidefinite.
By considering the characteristic polynomial of $M$: $\lambda^2- (\Tr M) \lambda + \det M=0$, this condition is equivalent to the two inequalities:
\begin{align} 
	0 \le  \Tr M = E-E_n+1,\text{ and } 0 \le \det M = E-E_n-|\Braket{\mathcal{A}_n}|^2.
\end{align}
The second inequality requires $E-E_n \ge 0$, and this condition is stronger than the first inequality ($0 \le  \Tr M $). Thus, the condition $ M  \succeq 0$ becomes
\begin{align} 
E-E_n \ge |\Braket{\mathcal{A}_n}|^2= \Braket{p}^2 + \left\{  \mu \Braket{e^x} -(h-n)  \right\}^2.
	\label{eq-ex-cond}
  \end{align}  
When $h-n \ge 0$, the minimum of the right hand side can be zero at $\Braket{p}=0$ and $\mu \Braket{e^x} = h-n$, and we obtain $E-E_n \ge 0$.
However, when $h-n \le 0$, since $\mu \Braket{e^x} \ge 0$, the minimum of the right hand side is $(h-n)^2$ at $\mu \Braket{e^x} = \Braket{p}= 0$, and the condition \eqref{eq-ex-cond} becomes
\begin{align} 
	E  \ge E_n +  \left(h-n\right)^2 = h^2, \quad (n \ge h),
	\label{eq-Morse-E-ge-h2}
	\end{align}
	where we have used $E_n=h^2-(h-n)^2$.

To summarize our discussions so far, if $E(E-E_1)\cdots(E-E_{n-1}) > 0$, we have the following condition on $E$.
\begin{align} 
	\begin{cases}
	E \ge E_n, \quad &(n \le h), \\
	 E \ge h^2, \quad &(n \ge h).
	\end{cases} 
	\label{eq-Morse-E-ge-h3}
\end{align}
(Note that $E_n=h^2$ when $n=h$.)
By taking $n=[h]'+1 (\ge h)$ in this result and combining it with Eq.~\eqref{eq-Morse-E-bootstrap}, we obtain Eq.~\eqref{eq-Morse-E-bootstrap-final}.
(In order to take $n=[h]'+1$, we need to take $K > [h]' +1$.)
This is the proof for Eq.~\eqref{eq-Morse-E-bootstrap-final}\footnote{In the region $E \ge h^2$, the energy eigenstates are non-normalizable and the assumption $\langle E | E \rangle =1$ imposed in Eq.~\eqref{bootstrap-XP} is violated.
However, even in this case, the bootstrap method yields the correct result \eqref{eq-Morse-E-ge-h3}. This can be explained as follows. For $E \ge h^2$, $\langle E | E \rangle$ diverges and the state $\Ket{E}$ does not belong to the Hilbert space.
However, we can formally regularize $\Ket{E}$ and treat $\langle E | E \rangle$ as a finite positive real number. 
Then we divide the bootstrap matrix \eqref{bootstrap} by $\langle E | E \rangle$ and make each component of the bootstrap matrix of the form $\Bra{E} O_m^\dagger O_n \Ket{E}/ \langle E | E \rangle$, which can be interpreted as the expectation value of $O_m^\dagger O_n$ evaluated by the formally regularized state $\Ket{E}$. 
Now, the normalization of the bootstrap matrix, where the (1,1) component is set to 1 as in Eq.~\eqref{bootstrap-XP}, is justified. 
In the Morse potential, the regularized expectation values of $\langle e^{mx} p^n \rangle$ ($m,n=0,1,\cdots$) are all finite even for the non-normalizable states, and we can perform the bootstrap analysis in the same way as in the normalizable case.
Then, the result \eqref{eq-Morse-E-ge-h3} is justified. In addition, the results of other observables, such as $\Braket{e^x}=0$ for $E \ge h^2$ in Fig.~\ref{fig-Morse-XP}, can also be understood consistently. (See also footnote \ref{ftnt-non-normalizable} for the definition of $\Braket{e^x}$.)
Similar things also happen in the Rosen-Morse potential and the hyperbolic Scarf potential cases discussed in Appendix \ref{app-SI-bootstrap}.
}.


\subsection{Derivation of $ \Braket{e^{x} }  $ in the bootstrap method}
\label{app-Morse-X}

We discuss the derivation of $ \Braket{e^{x} }  $ \eqref{eq-Morse-X} by using the bootstrap method. In Appendix \ref{app-Morse-off-diagonal}, we have studied the condition that the submatrix $\tilde{\mathcal M}$ \eqref{eq-sub-M} satisfies $\tilde{\mathcal M} \succeq 0 $. In particular, when $E(E-E_1)\cdots(E-E_{n-1}) > 0$, this condition leads to the inequality \eqref{eq-ex-cond}. If $n<h$, $E_m>E_{m-1}$ is held for ${}^\forall m < h$ and $E(E-E_1)\cdots(E-E_{n-1}) > 0$ is satisfied at $E=E_n$. Then, the condition \eqref{eq-ex-cond} is valid at $E=E_n$, and we obtain
  \begin{align}  
 \Braket{e^{x}}|_{E=E_n} = \frac{1}{\mu} (h-n), \quad \Braket{p}|_{E=E_n} =0 , \quad   (n<h) .
	\label{eq-ex}
\end{align}
This reproduces the exact result \eqref{eq-Morse-X} for the bound states.

On the other hand, for the states in the continuous spectrum ($E \ge h^2$), since the wave function spreads to $x \to -\infty$, $\Braket{e^{x}}$ should be zero.
However, $\Braket{e^{x}}=0$ for $E \ge h^2$ cannot be derived from the condition $\tilde{\mathcal M} \succeq 0 $ alone, and we need to consider a larger submatrix to show it. 
Although it seems a bit complicated to show it analytically, we can confirm that $\Braket{e^{x}}=0$ is obtained for $E \ge h^2$ by using the bootstrap method (See Fig.~\ref{fig-Morse-XP}).

Note that once we obtain $\Braket{e^{x}}$, we can determine other expectation values such as $\Braket{e^{mx} p^n} $ by using the relation \eqref{eq-recurrence-Morse}. Thus, the bootstrap method can derive various expectation values in the Morse potential.

\section{Bootstrapping shape invariant models}
\label{app-SI-bootstrap}

In this appendix, we study Rosen-Morse potentials and hyperbolic Scarf potentials, both of which are shape invariant. We will see that the bootstrap method reproduces the known analytic results for these potentials. The computation is similar to the Morse potential case, and we will not repeat the details.

\subsection{Bootstrapping Rosen-Morse potential}
\label{sec-RM}
\begin{figure}
	\centering
	\includegraphics[scale=0.5]{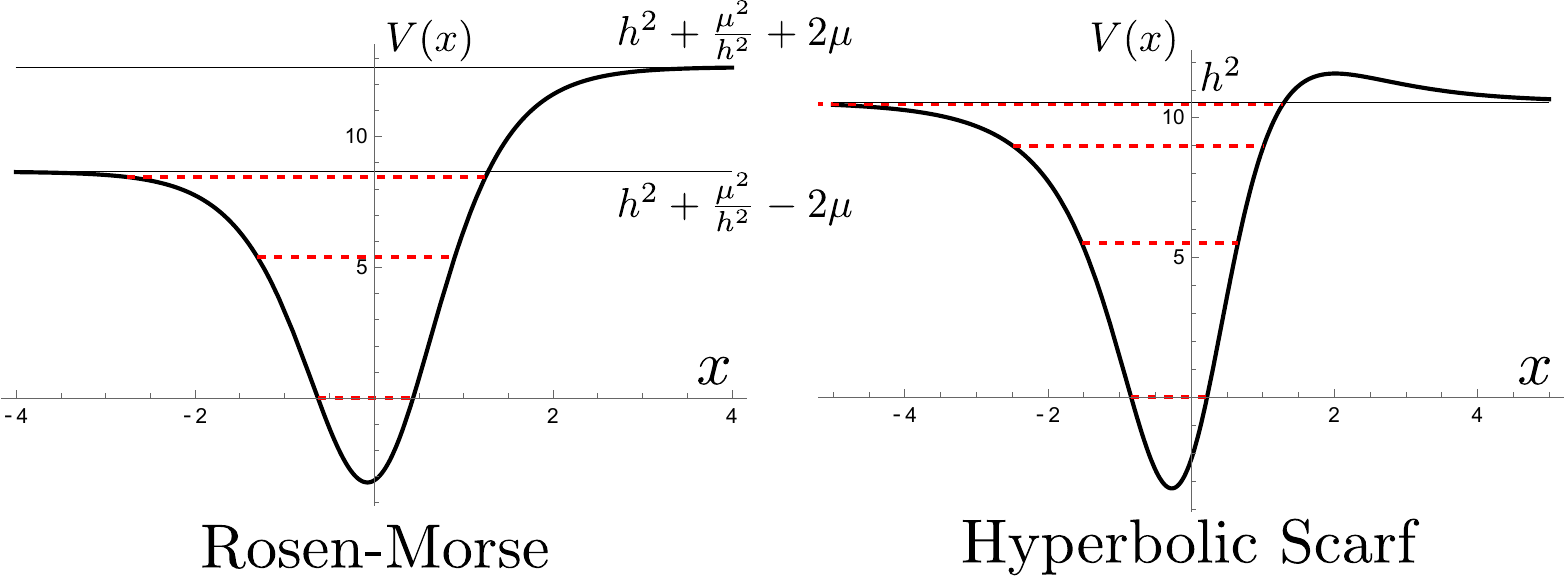}
	\caption{The Rosen-Morse potential \eqref{eq-RM-H} at $(h,\mu)=(13/4,1)$ and the hyperbolic Scarf potential at $(h,\mu)=(13/4,1)$. The red dashed lines represent the energy eigenvalues of the bound states \eqref{eq-RM-En} and \eqref{eq-HS-En}.}
\label{fig-RM-pot}
\end{figure}

We consider the Rosen-Morse potential \cite{PhysRev.42.210},
\begin{align} 
	\mathcal{H}=p^2+V(x), \quad V(x)= -\frac{h(h+1)}{\cosh^2 x}+  2 \mu \tanh x+ h^2 +\frac{\mu^2}{h^2},
\label{eq-RM-H}  
\end{align}
where $h > \sqrt{\mu}>0$. 
The potential is depicted in Fig.~\ref{fig-RM-pot}.
This model satisfies the shape invariant condition \eqref{eq-SI} with $(\lambda, \delta)=(h,-1)$ and the annihilation operator and the energy eigenvalue are given by
\begin{align} 
	\mathcal{A}_n & = ip + \frac{\mu}{h-n}+ (h-n) \tanh x, \nonumber \\
	 E_n &= h^2-(h-n)^2+\frac{\mu^2}{h^2}-\frac{\mu^2}{(h-n)^2}, \quad n=0,1,\cdots, [ h-\sqrt{\mu} ]'.
	\label{eq-RM-En}
\end{align}
For $E \ge h^2 + \frac{\mu^2}{h^2} -2\mu $, the continuous spectrum appears, and the energy eigenstates degenerate for $E \ge h^2 + \frac{\mu^2}{h^2} +2\mu $ as we can see in Fig.~\ref{fig-RM-pot}.

In addition, the following relations hold for $\langle \tanh x \rangle $ at the energy eigenstate:
\begin{align} 
&	\langle \tanh x \rangle|_{E=E_n}  =- \frac{\mu}{(h-n)^2}, \quad  n=0,1,\cdots, [ h-\sqrt{\mu} ]', \label{eq-RM-tanh}  \\
&	\langle \tanh x \rangle  =1, \quad h^2 + \frac{\mu^2}{h^2} -2\mu \le E < h^2 + \frac{\mu^2}{h^2} +2\mu , \label{eq-RM-tanh-C}   \\
	&-1 \le \langle \tanh x \rangle \le 1, \quad E \ge h^2 + \frac{\mu^2}{h^2} +2\mu. \label{eq-RM-tanh-D}
 \end{align}
 (See footnote \ref{ftnt-non-normalizable} for the definition of the expectation values of the non-normalizable states ($E \ge h^2 + \frac{\mu^2}{h^2} -2\mu $).)
Note that, for $E \ge h^2 + \frac{\mu^2}{h^2} +2\mu $, the expectation value $\langle \tanh x \rangle $ takes an undefined value $-1 \le \langle \tanh x \rangle \le 1 $ because of the degeneracy. (Here we have considered all possible expectation values of $\langle \tanh x \rangle $ for a given energy $E$, where the two non-normalizable energy eigenstates contribute to this.)
Similarly, the expectation value $\langle \tanh^2 x \rangle $ at the energy eigenstate is given by\footnote{\label{ftnt-referee}The authors would like to thank an anonymous referee for pointing out the formulae \eqref{eq-RM-tanh2}, \eqref{eq-HS-coshx2}, \eqref{eq-HS-sinhcoshx2}, \eqref{eq-HS-coshx1}, \eqref{eq-HS-coshx3} and \eqref{eq-HS-sinhcoshx1}.}
\begin{align} 
\langle  \tanh^2 x \rangle|_{E=E_n} & =\frac{1}{2h+1} \left(2n+1 + \frac{2\mu^2}{(h-n)^3}  \right), \quad  n=0,1,\cdots, [ h-\sqrt{\mu} ]', \label{eq-RM-tanh2} \\
\langle  \tanh^2 x \rangle & = 1, \quad E \ge h^2 + \frac{\mu^2}{h^2} -2\mu .
 \end{align}

\begin{figure}
	\centering
	\includegraphics[scale=0.7]{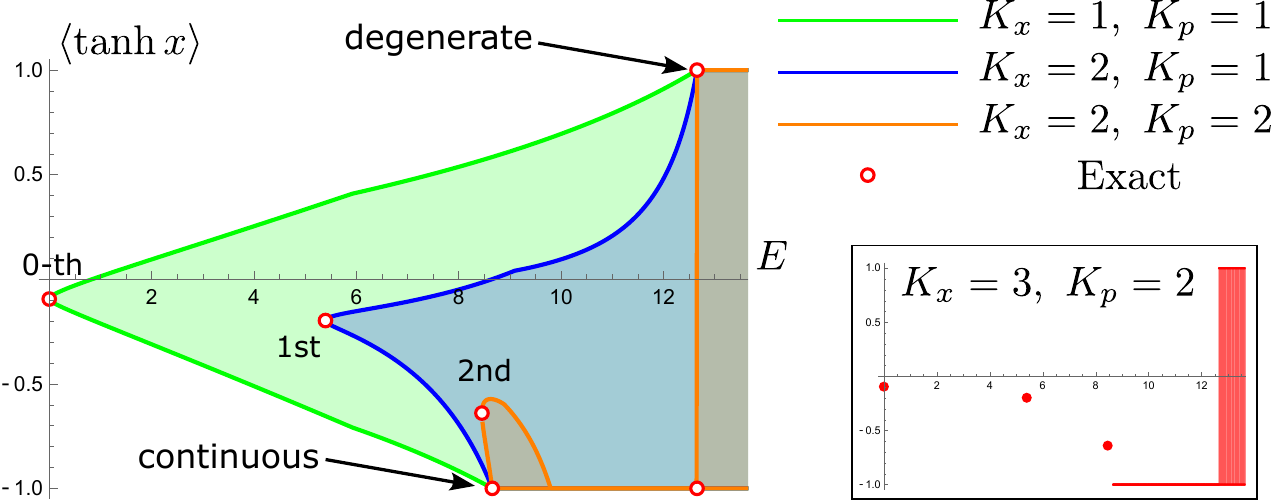}
	\caption{$E$ vs. $\langle \tanh x \rangle $ in the numerical bootstrap analysis of the Rosen-Morse potential \eqref{eq-RM-H} at $(h,\mu)=(13/4,1)$.
	We use the operator \eqref{operators-XP-RM} to construct the bootstrap matrix and determine the maximum and the minimum values of $\langle \tanh x \rangle $ at each $E$ by solving the linear programming.
	The inside of the curves and dots represent the allowed regions where the bootstrap matrix is positive-semidefinite. The red small circles represent the exact energy eigenstates \eqref{eq-RM-En} (the continuous spectrum is omitted). The exact solutions  \eqref{eq-RM-En} and \eqref{eq-RM-tanh} are reproduced at $(K_x,K_p)=(3,2)$. Since some allowed points are not visible because they overlap, see Table \ref{Table-RM} for the details.
	}
	\label{fig-RM}
\end{figure}

\begin{figure}
	\centering
	\includegraphics[scale=0.7]{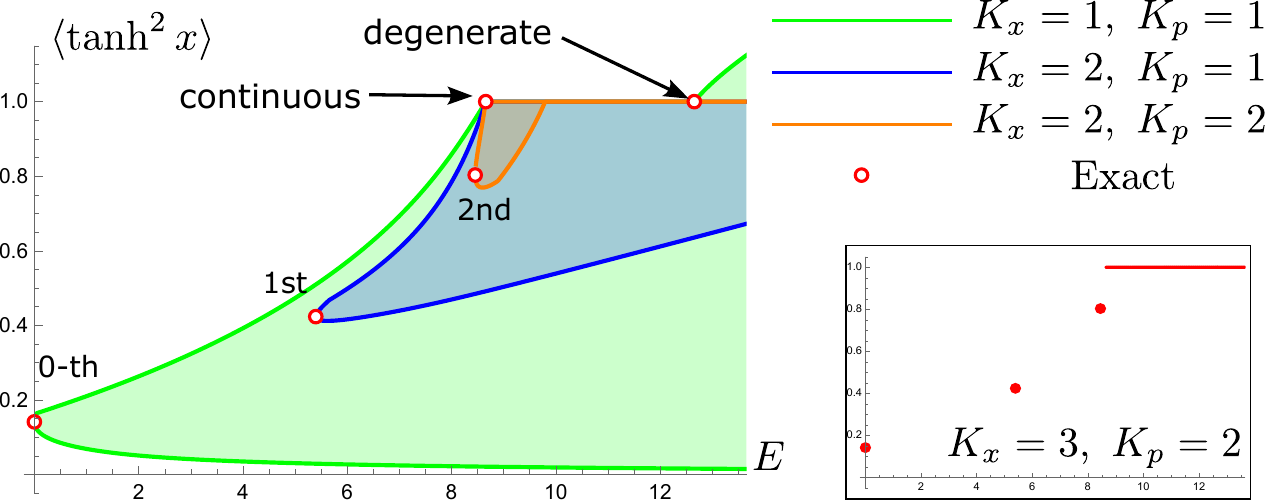}
	\caption{$E$ vs. $\langle \tanh^2 x \rangle $ in the Rosen-Morse potential \eqref{eq-RM-H} at $(h,\mu)=(13/4,1)$.
	We perform the same analysis as in Fig.~\ref{fig-RM} but we evaluate the maximum and the minimum values of $\langle \tanh^2 x \rangle $. }
	\label{fig-RM-tanh2}
\end{figure}

We apply the numerical bootstrap method to this model. We construct the bootstrap matrix by using the operators:
\begin{align}
	\tilde{O}:=  \sum_{m=0}^{K_x} \sum_{n=0}^{K_p} c_{mn} \left(\tanh x \right)^m p^n.
	\label{operators-XP-RM}
\end{align}
Through the constraints \eqref{HO=0} and \eqref{HO=EO}, the matrix elements of the bootstrap matrix can be expressed by $E$, $\langle \tanh x \rangle $, $\langle \tanh^2 x \rangle $, $\langle  p \rangle $ and $\langle (\tanh x )p \rangle $. We set $\langle  p \rangle $ = $\langle (\tanh x )p \rangle $ =0, which are satisfied for $E < h^2 + \frac{\mu^2}{h^2} +2\mu$, for simplicity.
Similarly to the Morse potential case, we numerically solve the linear programming and the results at $(h,\mu)=(13/4,1)$ are shown in Fig.~\ref{fig-RM} and \ref{fig-RM-tanh2}.
The result at $(K_x,K_p)=(3,2)$ are consistent with the known analytic results \eqref{eq-RM-En}, \eqref{eq-RM-tanh} and \eqref{eq-RM-tanh2}.
We can confirm that these are precisely agree with the analytic results by evaluating the characteristic polynomial of the bootstrap matrix.

Our study of the diagonal components of the bootstrap matrix $\mathcal{M}_\mathcal{A}$ in Sec.~\ref{sec-SI-bootstrap} ensures obtaining the bound state energies $E_n$. 
Also, the continuous spectrum $E \ge h^2 + \frac{\mu^2}{h^2} -2\mu $ and the expectation value $\langle \tanh x \rangle $ \eqref{eq-RM-tanh} and \eqref{eq-RM-tanh-C} can be derived through the contribution of the $(i,i+1)$ and $(i+1,i)$ components of the bootstrap matrix $\mathcal{M}_\mathcal{A}$ similar to the Morse potential case. These can be shown by using Eq.~\eqref{eq-sub-M} (See also footnote \ref{ftnt-general-SI}).

However, to show the agreement of $\langle \tanh^2 x \rangle $, we need to consider a larger submatrix of the bootstrap matrix $\mathcal{M}_\mathcal{A}$ but we do not pursue this in this paper.
Also, evaluating the observables in the continuum of the states in the region $E \ge h^2 + \frac{\mu^2}{h^2} +2\mu$ is interesting, but we do not pursue this either (Apparently, $\langle \tanh x \rangle $ and $\langle \tanh^2 x \rangle $ are correctly reproduced).

\begin{table}[t]
	\centering
	\begin{tabular}{|c||c|c|c|c|}
		\hline
		& $n= 0 $  & $n= 1 $ & $n= 2$  & continuous (no degeneracy)  \\ 
		\hline
		\hline
		$(K_x,K_p)=(1,1)$ &   \multicolumn{4}{c |}{$0 \le E $}\\
		\hline 
		$(K_x,K_p)=(2,1)$ &  0 & \multicolumn{3}{c |}{$ \frac{147763}{27378} \le E $}\\
		\hline 
		$(K_x,K_p)=(2,2)$ &  0 &  $\frac{147763}{27378}$ & \multicolumn{2}{c |}{$\frac{35721}{4225} \le E $}\\
		\hline 
		$(K_x,K_p)=(3,2)$ &  0 &  $\frac{147763}{27378}$ & $\frac{35721}{4225}$ & $\frac{23409}{2704} \le E < \frac{34225}{2704} $   \\
		\hline 
		\hline 
		exact result for $E_n$  & 0 & $\frac{147763}{27378}$ & $\frac{35721}{4225}$ & $\frac{23409}{2704} \le E < \frac{34225}{2704} $   \\
		\hline 
		$\Braket{\tanh x}|_{E=E_n}$ & $-\frac{16}{169}$ & $-\frac{16}{81}$ & $-\frac{16}{25}$ &  1   \\
		\hline 
		$\Braket{\tanh^2 x}|_{E=E_n}$ & $\frac{310}{2197}$ & $\frac{926}{2187}$ & $\frac{502}{625}$ &  1    \\
		\hline 
	\end{tabular}
	\caption{Energy spectrum of the Rosen-Morse potential \eqref{eq-RM-H} at $(h,\mu)=(13/4,1)$ via the bootstrap method. We evaluate the characteristic polynomial of the bootstrap matrix discussed in Sec.~\ref{sec-SDP-char} by using Mathematica and obtain the allowed points, which rigorously agree with the energy spectrum $E_n$ \eqref{eq-RM-En},  $\Braket{\tanh x}$ \eqref{eq-RM-tanh} and $\Braket{\tanh^2 x}$ \eqref{eq-RM-tanh2}. 
	We obtain consistent results from the numerical linear programming shown in Figs.~\ref{fig-RM} and \ref{fig-RM-tanh2}, but the allowed points are slightly smeared due to numerical error. 
	}
	\label{Table-RM}
\end{table}

\subsection{Bootstrapping hyperbolic Scarf potential}
\label{sec-HS}

We consider the hyperbolic Scarf potential \cite{JWDabrowska_1988, Levai:1989eaa, Alvarez-Castillo:2006uzq}
\begin{align} 
	\mathcal{H}=p^2+V(x), \quad V(x)= \frac{-h(h+1) + \mu^2 + \mu (2 h+1) \sinh x }{\cosh^2 x}+   h^2,
\label{eq-HS-H}  
\end{align}
where $h, \mu>0$. 
The potential is depicted in Fig.~\ref{fig-RM-pot}.
This model satisfies the shape invariant condition \eqref{eq-SI} with $(\lambda, \delta)=(h,-1)$ and the annihilation operator and the energy eigenvalue are given by
\begin{align} 
	\mathcal{A}_n = ip + \frac{\mu}{\cosh x}  + (h-n) \tanh x, \quad E_n= h^2-(h-n)^2, \quad n=0,1,\cdots, [ h]'.
	\label{eq-HS-En}
\end{align}
For $E \ge h^2 $, the continuous spectrum appears.
Also, the following relations hold at the energy eigenstate:
\begin{align} 
& \left. \Braket{ \frac{1}{\cosh^{2} x}  }\right|_{E=E_n}  = \frac{1}{2} \frac{(2h+1)(h-n)}{(h+\frac{1}{2})^2+\mu^2},  \label{eq-HS-coshx2}  \\
& \left. \Braket{ \frac{\sinh x}{\cosh^{2} x}  }\right|_{E=E_n}  = - \frac{\mu (h-n)}{(h+\frac{1}{2})^2+\mu^2}~ \left( = -\frac{2\mu}{2h+1} \left. \Braket{ \frac{1}{\cosh^{2} x}  }\right|_{E=E_n} \right)  \label{eq-HS-sinhcoshx2}  .
 \end{align}
Other quantities are given by more sophisticated formulae
\begin{align} 
&  \left. \Braket{ \frac{1}{\cosh x}} \right|_{E=E_n} \nonumber \\
&= (h-n) \frac{\Gamma(h-n+\frac{1}{2}+i\mu)\Gamma(h-n+\frac{1}{2}-i\mu)}{\Gamma(h+1+i\mu)\Gamma(h+1-i\mu)} \notag \\
  &\quad \times \sum_{k=0}^n \left[ \binom{n}{k} \frac{((2k-1)!!)^2}{2^{2k} k!} (2h-n-k+1)_k \prod_{j=0}^{n-k-1} \left( \left(h-n+\frac{1}{2}+j\right)^2 + \mu^2 \right) \right],
   \label{eq-HS-coshx1}
  \\ 
&  \left. \Braket{ \frac{1}{\cosh^{3} x} } \right|_{E=E_n} \nonumber \\
&= -\frac{1}{2} (h-n) \frac{\Gamma(h-n+\frac{1}{2}+i\mu)\Gamma(h-n+\frac{1}{2}-i\mu)}{\Gamma(h+2+i\mu)\Gamma(h+2-i\mu)} \notag \\
&\quad \times \sum_{k=0}^n \Biggl[ \binom{n}{k} \frac{(2k-1)!! (2k+1)!! }{ (2k-1) 2^{2k}(k+1)! } (2h-n-k+1)_k \notag \\
  &\quad \times \left( 2(k+1)h^2 + (2n+3-(4n+3)k)h + (n^2-1)(2k-1) \right) \notag \\
  &\quad \times \prod_{j=0}^{n-k-1} \left( \left(h-n+\frac{1}{2}+j\right)^2 + \mu^2 \right) \Biggr],
     \label{eq-HS-coshx3} \\
&  \left. \Braket{ \frac{\sinh x}{ \cosh x } } \right|_{E=E_n} \nonumber \\
 &= -\mu \frac{\Gamma(h-n+\frac{1}{2}+i\mu)\Gamma(h-n+\frac{1}{2}-i\mu)}{\Gamma(h+1+i\mu)\Gamma(h+1-i\mu)} \Biggl[ \prod_{j=0}^{n-1} \left(\left(h-n+\frac{1}{2}+j\right)^2+\mu^2\right) \notag \\
  &\quad + (2h+1) \sum_{k=1}^n \Biggl\{ \binom{n}{k} \frac{((2k-1)!!)^2}{2^{2k} k!} \sum_{l=0}^{\min(k-1,n-k)} (n-k-l+1)_l \frac{(k-l)_l}{(k+1)_l}  \notag \\
  &\quad \times (2h-n-k+2+l)_{k-1-l} \prod_{j=0}^{n-k-1} \left( \left(h-n+\frac{1}{2}+j\right)^2+\mu^2 \right) \Biggr\} \Biggr]
   \label{eq-HS-sinhcoshx1} 
 \end{align}
 with the conventions $\sum_{j=m}^{m-1}*=0$, $\prod_{j=m}^{m-1}*=1$ and $(-1)!!=1$.
 Here $(a)_n:=a(a+1)(a+2)\cdots (a+n-1)$ is the Pochhammer symbol.
For $E \ge h^2$, these quantities except $\braket{\sinh x/ \cosh x}$ become zero  because the wave function spreads to $x \to  \pm \infty$.
$\braket{\sinh x/ \cosh x}$ takes an undefined value $-1 \le \langle \sinh x/ \cosh x \rangle \le 1 $ because of the degeneracy.
We will compare these analytic formulae with the numerical bootstrap results below.

\begin{figure}
	\centering
	\includegraphics[scale=0.7]{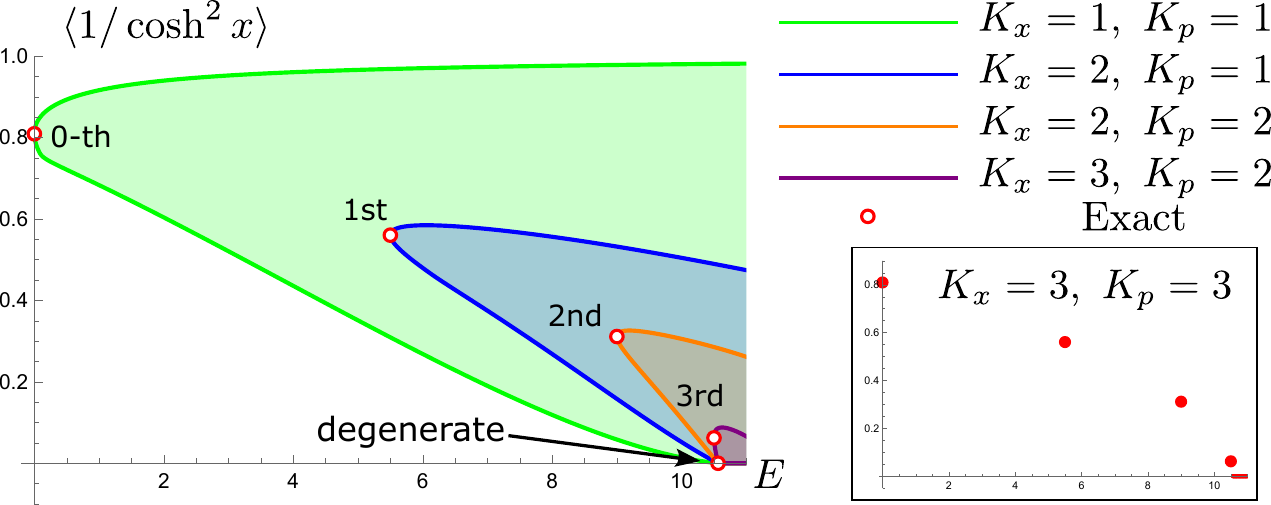}
	\caption{$E$ vs. $\langle 1/\cosh^2 x \rangle $ in the numerical bootstrap analysis of the hyperbolic Scarf potential \eqref{eq-HS-H} at $(h,\mu)=(13/4,1)$.
	We use the operator \eqref{operators-XP-HS} to construct the bootstrap matrix and determine the maximum and the minimum values of $\langle 1/\cosh^2 x  \rangle $ at each $E$ by solving the linear programming.
	The inside of the curves and dots represent the allowed regions where the bootstrap matrix is positive-semidefinite. The red small circles represent the energy eigenstates \eqref{eq-HS-En} and \eqref{eq-HS-coshx2} (the continuous spectrum is omitted). The exact results are (approximately) reproduced by the numerical bootstrap method at $(K_x,K_p)=(3,3)$.
	 Since some allowed points are not visible because they overlap, see Table \ref{Table-HS-E} for the details.
	}
	\label{fig-HS}
\end{figure}

\begin{figure}
	\centering
	\includegraphics[scale=0.7]{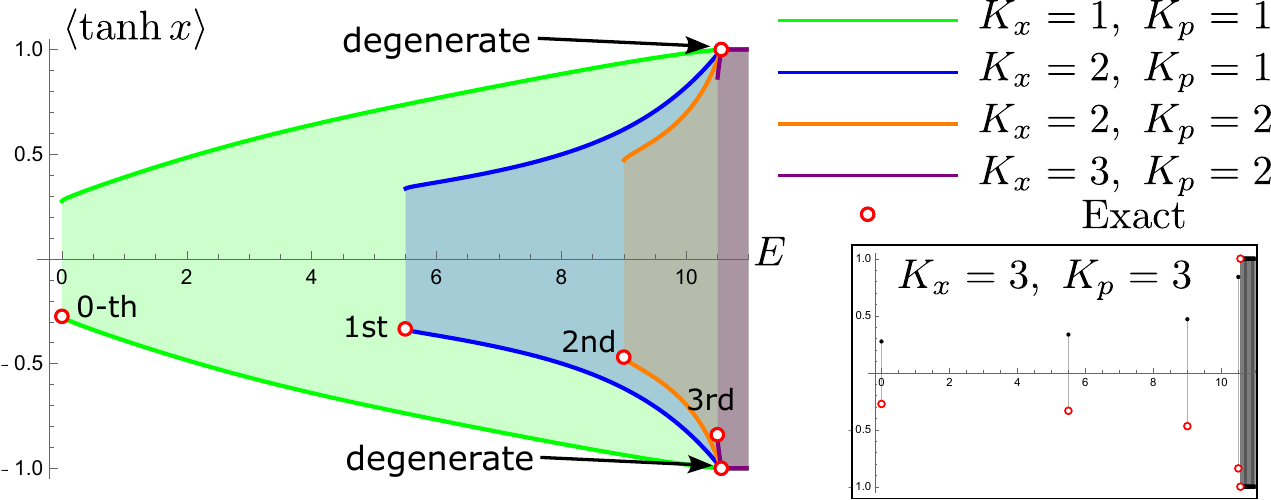}
	\caption{$E$ vs. $\langle \tanh x \rangle $ in the numerical bootstrap analysis of the hyperbolic Scarf potential \eqref{eq-HS-H} at $(h,\mu)=(13/4,1)$.
		We perform the same analysis as in Fig.~\ref{fig-HS} but we evaluate the maximum and the minimum values of $\langle \tanh x \rangle $. 
		 Since some allowed points are not visible because they overlap, see Table \ref{Table-HS-E} for the details.
	Although we obtain the energy eigenvalues consistent with Fig.~\ref{fig-HS}, the expectation value $\langle \tanh x \rangle $ is problematic. Firstly, the allowed regions are always symmetric as $\langle \tanh x \rangle \to -\langle \tanh x \rangle $, and they converge to lines rather than points. Secondly, although the boundaries of the allowed region are almost coincident with the exact values of $\langle \tanh x \rangle $ \eqref{eq-HS-sinhcoshx1}, they are not rigid boundaries. See Table \ref{Table-HS} for the details. 	
	}
	\label{fig-HS-tanh}
\end{figure}

\begin{figure}
	\centering
	\includegraphics[scale=0.55]{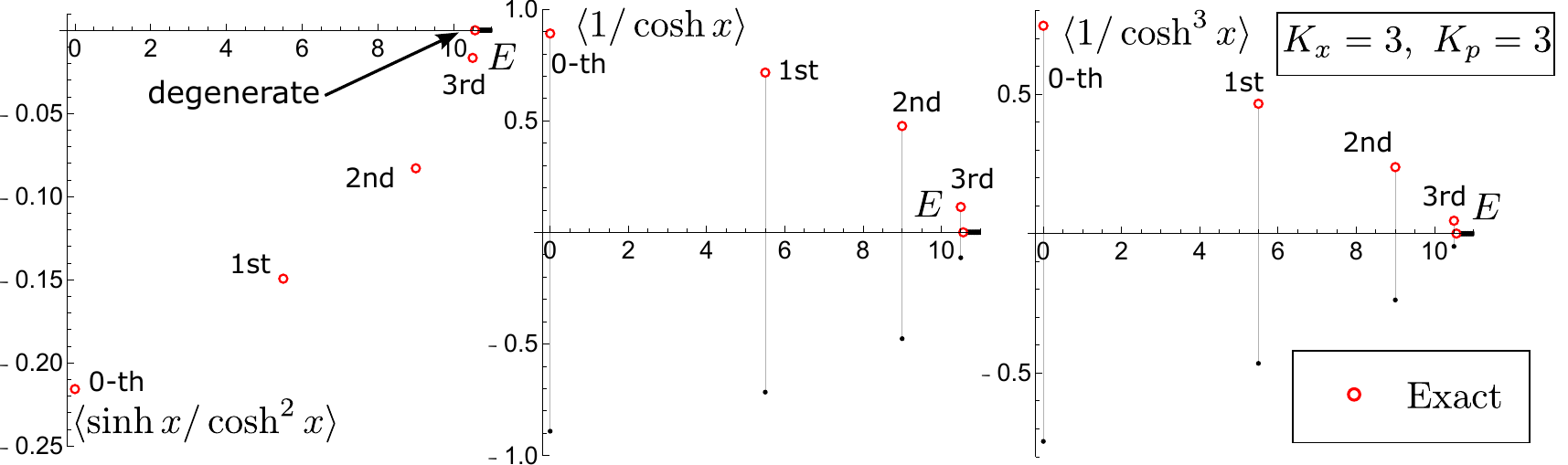}
	\caption{$E$ vs. $\langle \sinh x/ \cosh^2 x \rangle $, $\langle 1/ \cosh x \rangle$ and $\langle 1/ \cosh^3 x \rangle$ in the numerical bootstrap analysis of the hyperbolic Scarf potential \eqref{eq-HS-H} at $(h,\mu)=(13/4,1)$.
	We perform the numerical bootstrap analysis at $(K_x,K_p)=(3,3)$ and find the maximum and minimum values of $\langle \sinh x/ \cosh^2 x \rangle $, $\langle 1/ \cosh x \rangle$ and $\langle 1/ \cosh^3 x \rangle$ at each $E$.
The red small circles represent the exact energy eigenstates \eqref{eq-HS-En}, \eqref{eq-HS-sinhcoshx2}, \eqref{eq-HS-coshx1} and \eqref{eq-HS-coshx3} (the continuous spectrum is omitted).
The black dots and lines are the allowed regions obtained by the bootstrap method.
While we (approximately) obtain the exact results for $\langle \sinh x/ \cosh^2 x \rangle $, the results for $\langle 1/ \cosh x \rangle$ and $\langle 1/ \cosh^3 x \rangle$ are not exact, similar to the $\langle \tanh x \rangle$ case shown in Fig.~\ref{fig-HS-tanh}.}
\label{fig-HS-others}
\end{figure}

We apply the bootstrap method to this model. We construct the bootstrap matrix by using the operators:
\begin{align}
	\tilde{O}:=  \left( \sum_{m=0}^{K_x} \sum_{n=0}^{K_p}  c_{mn} \left( \cosh  x \right)^{-m} p^n \right)+ \left( \sum_{m=1}^{K_x} \sum_{n=0}^{K_p}  d_{mn} \sinh x \left( \cosh  x \right)^{-m}  p^n \right) .
	\label{operators-XP-HS}
\end{align}
Through the constraint \eqref{HO=0} and \eqref{HO=EO}, we can show that the matrix elements of the bootstrap matrix can be expressed by $E$, $\langle 1/ \cosh^m  x \rangle $ ($m=1,2,3$), $\langle \sinh x / \cosh^{m}  x \rangle $ ($m=1,2$), $\langle p \rangle $, $\langle (1/\cosh x) p \rangle $ and $\langle (\sinh x/\cosh x) p \rangle $.
We set $\langle p \rangle = \langle (1/\cosh x) p \rangle = \langle (\sinh x/\cosh x) p \rangle =0$, which are satisfied at the bound state, for simplicity. 
We numerically solve the linear programming and the results at $(h,\mu)=(13/4,1)$ are shown in Fig.~\ref{fig-HS}, \ref{fig-HS-tanh} and \ref{fig-HS-others}. See also Table \ref{Table-HS-E}.
Note that the characteristic polynomial analysis is difficult in this model\footnote{We have tried to obtain the exact results by solving Eqs.~\eqref{eq-bootstrap-char} and \eqref{eq-bootstrap-char-deriv} using the Mathematica function \texttt{Solve}, but it is difficult. 
If we set $\langle 1/ \cosh x \rangle  = \langle \sinh x/ \cosh x \rangle =  \langle 1/ \cosh^3 x \rangle =0$, which is consistent with the bootstrap analysis, we can find the exact solutions of $E$, $\langle 1/ \cosh^2 x \rangle$ and $ \langle \sinh x/ \cosh^2 x \rangle $ at $n=0$ and 1, but cannot obtain the solutions for $n \ge 2$. We presume that this is because the size of the bootstrap matrix is too large.
}. To confirm that we obtain the smeared allowed points, we investigate the \texttt{Tolerance} dependence discussed in Appendix \ref{app-exact-numerical} in Fig.~\ref{fig-Tol-HSTII}. The result strongly indicates that the obtained isolated allowed regions are smeared allowed points.

In this way, we obtain the energy eigenvalues $E_n$ and the expectation values  $\langle 1/ \cosh^2  x \rangle $ and $\langle \sinh x/  \cosh^{2}  x \rangle$  consistent with the exact results \eqref{eq-HS-En}, \eqref{eq-HS-coshx2} and \eqref{eq-HS-sinhcoshx2} at $(K_x,K_p)=(3,3)$.
However, we cannot obtain the values of  $\langle 1/ \cosh^m  x \rangle $ ($m=1,3$) and $\langle \sinh x  /\cosh  x \rangle = \langle \tanh  x \rangle$ through the bootstrap method. As shown in Figs.~\ref{fig-HS-tanh} and \ref{fig-HS-others}, the isolated allowed regions for these quantities converge to lines rather than points.

\begin{figure}
	\centering
	\includegraphics[scale=0.5]{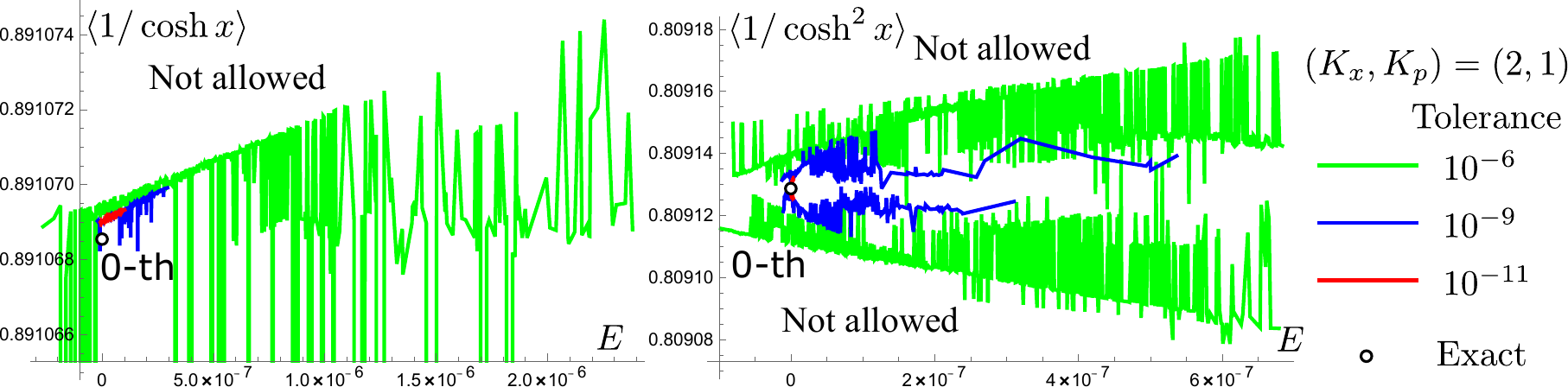}
	\caption{
	Enlarged view of the isolated allowed regions for the ground state ($n=0$) of the hyperbolic Scarf potential at $(K_x,K_p)=(2,1)$ and their \texttt{Tolerance} dependence.
	$E$ vs. $\langle 1/\cosh x \rangle $ and $\langle 1/ \cosh^2 x \rangle $ are shown. The curves represent the borders of the allowed regions obtained by the numerical linear programming at various \texttt{Tolerance} values.
	The sizes of the allowed regions significantly depend on the \texttt{Tolerance}, and the allowed regions converge to the exact values of $E$ and $\langle 1/ \cosh^2 x \rangle $ as the \texttt{Tolerance} decreases. This is a strong indication that the isolated allowed regions for the energy are smeared allowed points.
	As shown in Fig.~\ref{fig-HS-others}, the lower bound of the allowed region for $E$ vs. $\langle 1/\cosh x \rangle $ is in the negative $\langle 1/\cosh x \rangle $ region and is not presented in this figure. 
	}
	\label{fig-Tol-HSTII}
\end{figure}

\begin{table}[t]
	\centering
	\begin{tabular}{|c||c|c|c|c|c|}
		\hline
		& $n= 0 $  & $n= 1 $ & $n= 2$  & $n= 3$ & continuous \\ 
		\hline
		\hline
		$(K_x,K_p)=(1,1)$ &   \multicolumn{5}{c |}{$0 \le E $}\\
		\hline 
		$(K_x,K_p)=(2,1)$ &  0 & \multicolumn{4}{c |}{$11/2 \le E $}\\
		\hline 
		$(K_x,K_p)=(2,2)$ &  0 &  11/2 & \multicolumn{3}{c |}{$9 \le E $}\\
		\hline 
		$(K_x,K_p)=(3,2)$ &  0 &  11/2 & 9 & \multicolumn{2}{c |}{$21/2 \le E $}\\
		\hline 
		$(K_x,K_p)=(3,3)$ &  0 &  11/2 & 9 &  21/2  & $169/16 \le E$ \\
		\hline 
		\hline 
		exact result for $E_n$~\eqref{eq-HS-En}  & 0 & 11/2 & 9 &  21/2 &  $ 169/16 \le E$   \\
		\hline 
		$\langle 1/ \cosh^2 x \rangle |_{E=E_n}  ~\eqref{eq-HS-coshx2}  $ & $195/241$ & $135/241$ & $75/241$ & $15/241$ &  0   \\
		\hline 
		$\langle \sinh x/ \cosh^2 x \rangle |_{E=E_n} ~\eqref{eq-HS-sinhcoshx2} $  & $-52/241$ & $-36/241$ & $-20/241$ & $-4/241$ & 0  \\
		\hline 
	\end{tabular}
	\caption{Energy spectrum of the hyperbolic Scarf potential \eqref{eq-HS-H}  at $(h,\mu)=(13/4,1)$ via the bootstrap method. 
	The (smeared) allowed points obtained through the numerical linear programming analysis shown in Figs.~\ref{fig-HS}, \ref{fig-HS-tanh} and \ref{fig-HS-others} converge to the values of this table as the
	 \texttt{Tolerance} decreases (See Fig.~\ref{fig-Tol-HSTII}). Correspondingly, the values of $\langle 1/ \cosh^2 x \rangle $ and $\langle \sinh x/ \cosh^2 x \rangle $ at each energy eigenstates converge to the exact results \eqref{eq-HS-coshx2} and \eqref{eq-HS-sinhcoshx2}.
	 The values of $\langle 1/ \cosh x \rangle $ are discussed in Table \ref{Table-HS}. (We do not show the values of $\langle 1/ \cosh^3 x \rangle $ and $\langle \sinh x/ \cosh x \rangle $ because they are qualitatively similar to $\langle 1/ \cosh x \rangle $.)
	}
	\label{Table-HS-E}
\end{table}

The reason that we cannot obtain these quantities can be clarified by investigating the bootstrap matrix $\mathcal{M}_\mathcal{A}$ \eqref{eq-SI-M} constructed from the annihilation operator $\mathcal{A}_n$ \eqref{eq-HS-En}. 
The bootstrap matrix $\mathcal{M}_\mathcal{A}$ ($K=3$ and $(h,\mu)=(13/4,1)$) at $E=E_0=0$ becomes 
\begin{align} 
&	\mathcal{M}_\mathcal{A} |_{E=0}= \nonumber \\
&\left(
\begin{array}{ccccc}
 1 & c_1+\frac{13}{4} s_1 & 13s_2-\frac{149 c_2}{8}+\frac{143}{8} & \frac{6441}{82} c_1+\frac{1287}{16} s_1-\frac{10553}{164} c_3 & \frac{5331}{16}s_2-\frac{8291 c_2}{32}+\frac{9009}{32} \\
  & 0 & \frac{51}{41} c_1-\frac{61}{41} c_3 & 0 & \frac{51561 }{3977}c_1-\frac{61671 }{3977} c_3 \\
  &  & 0 & \frac{73810 }{3977}c_3-\frac{61710 c_1}{3977} & 0 \\
  & \text{h.c.}  & & 0 & \frac{84656583 }{326114}c_1-\frac{101255913 }{326114}c_3 \\
  &  &  & & 0 \\
\end{array}
\right)  ,
 \end{align}
where $c_m:= \langle 1/ \cosh^m x \rangle$ ($m=1,2,3$) and $s_m:= \langle \sinh x/ \cosh^m x \rangle$ ($m=1,2$).
This matrix is positive semidefinite if and only if the following conditions are satisfied:
\begin{align}
& \Braket{ \frac{1}{ \cosh^2 x} } = \frac{195}{241}, \quad \Braket{ \frac{\sinh x}{ \cosh^2 x} } = -\frac{52}{241},  \label{eq-HS-coshx2-E=0}\\
& \Braket{ \frac{1}{ \cosh x} } = \frac{61}{51}  \Braket{ \frac{1}{ \cosh^3 x} }  = - \frac{13}{4}  \Braket{ \frac{\sinh x}{ \cosh x} } ~~ \left(  = - \frac{13}{4}  \Braket{ \tanh x }  \right).
\label{eq-HS-others-E=0}
\end{align}
Eqs.~\eqref{eq-HS-coshx2-E=0} agree with the exact values \eqref{eq-HS-coshx2} and \eqref{eq-HS-sinhcoshx2} at $n=0$ (See also Table \ref{Table-HS-E}), and this explains why the bootstrap method can reproduce the exact values of $E_n$, $\langle 1/ \cosh^2 x \rangle $ and $\langle \sinh x/ \cosh^2 x \rangle $. 
However, $\langle 1/ \cosh^m  x \rangle $ ($m=1,3$) and $\langle \sinh x  \cosh^{-1}  x \rangle$ are undetermined. 
Note that any additional constraint for these expectation values do not appear from the bootstrap matrix $\mathcal{M}_\mathcal{A} |_{E=0}$ for $K \ge 4$.

To evaluate these undetermined quantities, we substitute the obtained relation \eqref{eq-HS-coshx2-E=0} and \eqref{eq-HS-others-E=0} into the bootstrap matrix $\mathcal{M}$ constructed from the operators \eqref{operators-XP-HS} at $E=E_0=0$.
The free parameter of this bootstrap matrix is only $\langle 1/ \cosh x \rangle $ through the relation \eqref{eq-HS-others-E=0}, and we can easily find the allowed region of the bootstrap matrix $\mathcal{M}$.
The result is shown in Table \ref{Table-HS}.
There are two important points in this table.
Firstly, the boundaries of the allowed regions depend on $K_x$ and $K_p$.
Thus, they are not rigid boundaries discussed in Sec.~\ref{sec-boundaries}, and the bootstrap method cannot derive the exact value of $\langle 1/ \cosh x \rangle $ (and $\braket{1/ \cosh^3 x}$ and $\braket{\tanh x}$).
Secondly, the allowed region is symmetric under the reflection $\langle 1/ \cosh x \rangle \to -\langle 1/ \cosh x \rangle $.
Thus, although $\langle 1/ \cosh x \rangle$ is a positive quantity, negative value is allowed in the bootstrap method.
Why does this happen?
The answer is that the bootstrap method in this model does not distinguish the expectation value of $\langle 1/ \cosh x \rangle$ and $-\langle 1/ \cosh x \rangle$. 
To be more precise, the constraints \eqref{HO=0} and \eqref{HO=EO} are invariant under the simultaneous reflection transformation $(p, 1/ \cosh x ,
  \tanh x) $ to $(-p, -1/ \cosh x ,  -\tanh x )$, since the Hamiltonian \eqref{eq-HS-H} is invariant under this transformation. (This is not the parity transformation $(p,x) \to (-p,-x)$.)
Since the bootstrap analysis uses only these constraints, it cannot distinguish the expectation values of $\langle 1/ \cosh x \rangle$ and $-\langle 1/ \cosh x \rangle$.
As a result, the obtained allowed region is symmetric under the reflection $\langle 1/ \cosh x \rangle \to -\langle 1/ \cosh x \rangle $.
Similar things occur to the results of $\langle 1/ \cosh^3 x \rangle$ and $\braket{\tanh x}$ too (See Fig.~\ref{fig-HS-tanh} and \ref{fig-HS-others})\footnote{A similar thing occurs in the harmonic and anharmonic oscillator cases. The Hamiltonian \eqref{H-HO} and \eqref{H-AHO} are symmetric under the parity $(p,x) \to (-p,-x)$, and the bootstrap analysis does not distinguish $(p,x)$ and $(-p,-x)$. However, the expectation values of the parity odd operators are zero and this issue does not appear there.} \footnote{
The situation may change if additional operators are included in the operator $\tilde{O}$ in Eq.~\eqref{operators-XP-HS}. This issue will be addressed in future work.	
}.

Although the bootstrap method cannot derive the exact values of these quantities, it still can provide the (approximated) exact energy eigenvalues as shown in Fig.~\ref{fig-Tol-HSTII}. 
Thus, the bootstrap analysis indicates the solvability of this model.

\begin{table}
	\centering
	\begin{tabular}{|c||c|}
		\hline
		&  $\langle 1/ \cosh x \rangle $  ($n=0$)      \\ 
		\hline
		\hline
$(K_x,K_p)=(1,1)$  &   $ -0.89110076499 \le \langle 1/ \cosh x \rangle  \le 0.89110076499  $   \\
\hline 
$(K_x,K_p)=(2,1)$  &   $-0.89106899288 \le \langle 1/ \cosh x \rangle  \le 0.89106899288  $   \\
\hline 
$(K_x,K_p)=(2,2)$  &   $ -0.89106855466 \le \langle 1/ \cosh x \rangle  \le 0.89106855466 $   \\
\hline 
$(K_x,K_p)=(3,2)$  &   $ -0.89106854679 \le \langle 1/ \cosh x \rangle  \le 0.89106854679  $   \\
\hline 
$(K_x,K_p)=(3,3)$  &   $ -0.89106854663 \le \langle 1/ \cosh x \rangle  \le 0.89106854663$   \\
\hline 
		\hline 
Exact ~\eqref{eq-HS-coshx1} &  $\langle 1/ \cosh x \rangle =0.89106854663\ldots$   \\
		\hline 
	\end{tabular}
	\caption{
		The allowed value of $\langle 1/ \cosh x \rangle $ at the ground state ($n=0$) in the hyperbolic Scarf potential \eqref{eq-HS-H} at $(h,\mu)=(13/4,1)$ in the bootstrap method.
		We use the bootstrap matrix constructed from the operators \eqref{operators-XP-HS}, and the relations \eqref{eq-HS-coshx2-E=0}, \eqref{eq-HS-others-E=0} and $E=0$ are substituted.
		The result is symmetric under the reflection $\langle 1/ \cosh x \rangle \to -\langle 1/ \cosh x \rangle $, and the allowed region does not converges to a point.
		The upper bound of the allowed region asymptotically approaches the exact result. This behavior is similar to the anharmonic oscillator case in Table \ref{Table-AHO-XP}, and this is not a rigid boundary discussed in Sec.~\ref{sec-boundaries}.
		}
	\label{Table-HS}
\end{table}

\section{Krein-Adler transformation and bootstrap method}
\label{app-DT}

We can generate new solvable systems from a seed solvable system using the Krein-Adler transformation \cite{zbMATH03242479, zbMATH00916231}. In this appendix, we will show that the bootstrap method can derive the exact enegy eigenvalues for the generated systems, if the original system has shape invariance.\\

First, we briefly review the construction of solvable systems using the Krein-Adler transformation \cite{zbMATH03242479, zbMATH00916231}, which is related to the works done by Darboux \cite{darboux1882proposition} and Crum \cite{10.1093/qmath/6.1.121, 1999physics...8019C}.
Suppose that a solvable system has a Hamiltonian $H^{(0)}$, the eigenstate $\ket{n}^{(0)}$, and the energy eigenvalue $E_n$,
\begin{align} 
	H^{(0)} \ket{n}^{(0)} = E_n \ket{n}^{(0)}, \quad n=0,1,2,\cdots.
	\label{eq-DT-H0}
\end{align}
For simplicity, we assume $E_n<E_{n+1}$ for ${}^\forall n$.
Here we formally introduce the annihilation operator $B_{m}^{(0)}$ for the $m$-th state $\ket{m}^{(0)}$, and rewrite the Hamiltonian as,
\begin{align} 
	H^{(0)}=B_{m}^{(0)\dagger} B_{m}^{(0)} + E_m, \quad  B_{m}^{(0)} \ket{m}^{(0)}=0.
	\label{eq-DT-H0-2}
\end{align}
This may be a formal expression, because $B_{m}^{(0)}$ is singular in general\footnote{If $ B_{m}^{(0)}$ is non-singular, $0 \le \bra{E_n} B_{m}^{(0)\dagger } B_{m}^{(0)} \ket{E_n} =E_{n}-E_{m} $ is required for ${}^\forall n$. This is not possible unless $\ket{m}^{(0)}$ is the ground state $(m=0)$, and indicates that $ B_{m}^{(0)}$ ($m \ge 1$) is singular. Note that this relation can be regarded as a bootstrap analysis with $O_m=B_{m}^{(0)}$ in Eq.~\eqref{ops-sample}, and shows that the bootstrap analysis with singular operators is problematic. }.
We then define the new Hamiltonian $H^{(1)}$ as
\begin{align} 
	H^{(1)}:=B_{m}^{(0)} B_{m}^{(0)\dagger} + E_m.
	\label{eq-DT-H1}
\end{align}
Now we formally obtain the energy eigenstates of $H^{(1)}$ from the original eigenstates as
\begin{align} 
	\ket{n}^{(1)}  \propto B_{m}^{(0)} \ket{n}^{(0)}, \quad H^{(1)} \ket{n}^{(1)}=E_n \ket{n}^{(1)},
\end{align}
where we can easily show the second equation by using Eqs.~\eqref{eq-DT-H0}, \eqref{eq-DT-H0-2} and \eqref{eq-DT-H1}.
Note that the $m$-th eigenstate has been deleted in this new system because $B_{m}^{(0)} \ket{m}^{(0)}=0$.
Thus, we obtain the new solvable system with the same energy eigenvalue $\{E_n\}$ as the original system except $E_m$. 
By repeating this procedure $M$ times, we can delete $M$ states from the original system. 
We denote the deleted states as $(m_1,m_2,\cdots,m_M)$.
Krein and Adler found that the obtained system is non-singular if the following condition is satisfied \cite{zbMATH03242479, zbMATH00916231}
\begin{align} 
  \prod_{j=1}^{M} (m-m_j) \ge 0, \quad ({}^\forall m \in \mathbf{Z}_{\ge 0}).
\label{eq-KA-cond}
\end{align}
This map from the original solvable system to the new solvable system is called the Krein-Adler transformation. The energy spectrum of the obtained system is equivalent to the original system $\{E_n\}$ except the deleted states $(m_1,m_2,\cdots,m_M)$.\\

Note that even if the original solvable system is shape invariant, the new system does not satisfy shape invariance in general. We now show that the bootstrap method can derive the energy eigenvalues of the new system in this case. For simplicity, we consider the $M=2$ case (we can easily generalize this to $M \neq 2$).
The condition \eqref{eq-KA-cond} for $M=2$ requires that these two states have to be consecutive, say $m$ and $m+1$. Hence, we consider the following system:
\begin{align} 
	H^{(2)}:=&B_{m+1}^{(1)} B_{m+1}^{(1)\dagger} + E_{m+1}, \nonumber \\
		H^{(1)}=&B_{m+1}^{(1)\dagger} B_{m+1}^{(1)} + E_{m+1}=  B_{m}^{(0)} B_{m}^{(0)\dagger} + E_m, 
	 \quad 	H^{(0)}=  B_{m}^{(0)\dagger} B_{m}^{(0)} + E_m,
	\nonumber \\
		& \ket{n}^{(2)} \propto B_{m+1}^{(1)} B_{m}^{(0)} \ket{n}^{(0)}, \quad B_{m+1}^{(1)} B_{m}^{(0)} \ket{m+1}^{(0)}=0, \quad B_{m}^{(0)} \ket{m}^{(0)}=0.
		\label{eq-DT-H2}
\end{align}
We are interested in the case where the original system is shape invariant, and we assume that the original Hamiltonian $H^{(0)}$  and the energy eigenstates $\ket{n}^{(0)}$ are expressed as
\begin{align} 
	H^{(0)}= \mathcal{A}_0^{\dagger} \mathcal{A}_0+E_0, \quad H^{(0)} \ket{n}^{(0)}=E_n\ket{n}^{(0)},
	\quad \ket{n}^{(0)} \propto	\mathcal{A}_0^{\dagger}\mathcal{A}_1^{\dagger}\cdots\mathcal{A}_{n-1}^{\dagger}\ket{0}_n^{(0)}, 
\end{align}
where $\mathcal{A}_n$ satisfies the shape invariant condition \eqref{eq-SI}.

We evaluate the energy eigenvalues of $H^{(2)}$ by using the bootstrap method. When we studied the bootstrap method in the shape invariant system in Sec.~\ref{sec-SI-bootstrap}, we found that constructing the bootstrap matrix by using the operator $\mathcal{A}_{n} \mathcal{A}_{n-1} \cdots \mathcal{A}_1 \mathcal{A}_0$, which annihilates the state $\ket{n}^{(0)}$, is useful.
Here we can easily show that the state $\ket{n}^{(2)}$ \eqref{eq-DT-H2} is annihilated as
\begin{align} 
	\mathcal{A}_{n} \mathcal{A}_{n-1} \cdots \mathcal{A}_1 \mathcal{A}_0  B_{m}^{(0)\dagger}  B_{m+1}^{(1)\dagger}	\ket{n}^{(2)}=0.
\end{align}
This motivates us to construct the bootstrap matrix by using the operator
\begin{align}
	\tilde{O}_\mathcal{A} := c_0 I + \sum_{n=0}^{K} c_{n} \left(  \mathcal{A}_{n-1}
	\mathcal{A}_{n-2} \cdots \mathcal{A}_1 \mathcal{A}_0 B_{m}^{(0)\dagger}  B_{m+1}^{(1)\dagger} \right).
	\label{eq-DT-O_A}
	\end{align}
Then, we can perform almost the same analysis as in the shape invariant case.
The diagonal component of the bootstrap matrix constructed from this operator can be evaluated as
\begin{align} 
& \Braket{ B_{m+1}^{(1)} B_{m}^{(0)}   \mathcal{A}_0^{\dagger}\mathcal{A}_1^{\dagger}\cdots \mathcal{A}_{n-1}^{\dagger} \mathcal{A}_{n}^{\dagger}\mathcal{A}_{n}  \mathcal{A}_{n-1}\cdots \mathcal{A}_1 \mathcal{A}_0 B_{m}^{(0)\dagger}  B_{m+1}^{(1)\dagger}} \nonumber \\
& =
(E-E_{n})(E-E_{n-1})\cdots(E-E_0) \Braket{ B_{m+1}^{(1)} B_{m}^{(0)}     B_{m}^{(0)\dagger}  B_{m+1}^{(1)\dagger}} \nonumber \\
& =
(E-E_{n})(E-E_{n-1})\cdots(E-E_0) (E-E_{m})(E-E_{m+1}) ,
\end{align}
where we have used the relation $\bra{E} H^{(2)}=E \bra{E}$ and $B_{m+1}^{(1)} B_{m}^{(0)} H^{(0)}= H^{(2)} B_{m+1}^{(1)} B_{m}^{(0)} $, which can be derived from Eq.~\eqref{eq-DT-H2}.
Since all the diagonal components of the bootstrap matrix have to be non-negative, we obtain the same condition to that of the shape invariant case \eqref{eq-SI-E-bootstrap}, and only $E=E_n$ ($n=0,1,\cdots$) is allowed. 
Thus, the bootstrap method reproduces the exact energy eigenvalues.
Note, however, that the energies $E_m$ and $E_{m+1}$, which are deleted by the Krein-Adler transformation, are not excluded in this analysis, where only the diagonal components of the bootstrap matrix are considered. These states might be removed by evaluating the entire bootstrap matrix.
Although, this point is subtle, our analysis in this appendix shows that the bootstrap method can determine the solvability of the system, which is generated by the Krein-Adler transformation.
Of course, through the relation \eqref{eq-O-A-SI}, the same result can be obtained without using the annihilation operators in the bootstrap matrix.

{\normalsize 
\bibliographystyle{utphys}
\bibliography{QM} }

\end{document}